\documentclass[11pt,a4paper]{article}
\pdfoutput=1

\usepackage{jheppub}

\usepackage{amsfonts,amssymb}
\usepackage[mathscr]{eucal}
\usepackage{amsmath}
\usepackage{graphicx}

\newcommand{\be}{\begin{equation}}
\newcommand{\ee}{\end{equation}}
\newcommand{\bea}{\begin{eqnarray}}
\newcommand{\eea}{\end{eqnarray}}
\newcommand{\ba}{\begin{eqnarray}}
\newcommand{\ea}{\end{eqnarray}}

\newcommand{\beq}{\begin{equation}}
\newcommand{\eeq}{\end{equation}}
\newcommand{\beqa}{\begin{eqnarray}}
\newcommand{\eeqa}{\end{eqnarray}}
\newcommand{\beqar}{\begin{eqnarray*}}
\newcommand{\eeqar}{\end{eqnarray*}}

\title{Holography, Hydrodynamization\\ and Heavy-Ion Collisions}

\author{Michal P.\ Heller$^{*}$\note[*]{On leave from: \emph{National Centre for Nuclear Research, Ho{\.z}a 69, 00-681 Warsaw, Poland}.}}

 \affiliation{Perimeter Institute for Theoretical Physics,\\
 31 Caroline Street North, Waterloo, Ontario N2L 2Y5, Canada}

\emailAdd{mheller@perimeterinstitute.ca}

\abstract{
In the course of the past several years holography has emerged as an ab initio tool in exploring strongly-time-dependent phenomena in gauge theories. These lecture notes overview recent developments in this area driven by phenomenological questions concerning applicability of hydrodynamics (hydrodynamization) under extreme conditions occurring in ultrarelativistic heavy-ion collisions at RHIC and LHC. The topics include hydrodynamization time scale, holographic collisions, as well as hydrodynamization from the point of view of the asymptotic character of the hydrodynamic gradient expansion. The emphasis is put on concepts rather than calculational techniques and a particular attention is devoted to present these developments in the context of the most recent advances and some of the open problems.
}
\notoc

\begin{document}  

\maketitle

\newpage

\section{Introduction and motivation}

Arguably one of the most interdisciplinary trends rooted in the theoretical high energy physics of the last 15 years has been efforts to shed light on time-dependent processes in strongly-coupled gauge theories, in particular ${\cal N} = 4$ super Yang-Mills (SYM) theory, using holography~\cite{Maldacena:1997re,Gubser:1998bc,Witten:1998qj} and gravity in anti-de Sitter (AdS) spacetimes. These fascinating developments were primarily motivated by the newly discovered quark-gluon plasma (QGP) in ultrarelativistic heavy-ion collisions (HICs) at RHIC~\cite{Gyulassy:2004zy} and subsequently explored also at the LHC~\cite{Muller:2012zq}. The QGP is a phase of quantum chromodynamics (QCD) in which the fundamental constituents of atomic nuclei, quarks and gluons, are deconfined but not necessarily weakly-interacting, see Ref.~\cite{Shuryak:1988ck} for a pedagogical exposition of these ideas.

The key HIC phenomenological notion are features of the produced particles univocally interpreted as signs of an underlying collective behaviour at least in some of the collisions. Interested reader is invited to consult Sec. 2 of Ref.~\cite{Wiedemann:2012py} for a brief yet quite comprehensive overview of this topic. In the context of these lectures, the collective behaviour is taken to simply mean that nuclear physicists are able to successfully fit the experimental signal using, as an intermediate step, equations of motion for the expectation value of the energy-momentum tensor $\langle T^{\mu \nu}\rangle$ of the created matter of the form
\be
\label{eq.TmunuMAIN}
\langle T^{\mu \nu} \rangle = {\cal E}(T) u^{\mu}u^{\nu} + P(T) \left( \eta^{\mu \nu} + u^{\mu} u^{\nu} \right) + \Pi^{\mu \nu}.
\ee
The equations of motion that allow to reproduce the observed particle spectra, referred to throughout the text as to various hydrodynamic equations, are conservation of the energy-momentum tensor \eqref{eq.TmunuMAIN} and a class of simple\footnote{Otherwise one looses the predictive power by having too many parameters to fit. For the relations used in solving the actual initial value problem for relativistic viscous fluids, see, e.g., Ref.~\cite{Baier:2007ix}.} relations for the absent-in-equilibrium contributions collectively denoted by $\Pi^{\mu \nu}$. 

The acclaimed discovery of QGP lies, in particular, in the successful use of the above model with ${\cal E}(T)$ and $P(T)$ being the equilibrium energy density and pressure (both are taken to be functions of temperature $T$) of the QGP phase obtainable from QCD using well-established theoretical approaches. As it also turned out, 
\be
\label{eq.hydro0}
\Pi^{\mu \nu} = 0
\ee
corresponding to perfect (i.e. without entropy production) fluid behaviour was disfavoured by the data creating an invaluable opportunity to gain insights into the microscopic physics of the experimentally accessible QGP. The reason for it is that the next-to-the-simplest result for the form of $\Pi^{\mu \nu}$, the so-called first (i.e. containing at most one derivative of $T$ and $u^{\alpha}$) order hydrodynamics given by\footnote{The term $\sigma^{\mu \nu}$ in Eq.~\eqref{eq.hydro1}, the so-called shear tensor, is given by
\be
\label{eq.sigma}
\sigma^{\mu \nu} = \Delta^{\mu \alpha} \Delta^{\nu \beta} \left( \nabla_{\alpha} u_{\beta} + \nabla_{\beta} u_{\alpha} \right) - \frac{2}{3} \Delta^{\mu \nu} \nabla_{\alpha} u^{\alpha}
\ee
and is traceless and transversal, i.e. $u_{\mu} \sigma^{\mu \nu} = 0$. In fact, we demand the whole $\Pi^{\mu \nu}$ to be transversal as a definition of velocity (this is the so-called Landau frame condition). The role of $\sigma^{\mu \nu}$ in the equations of motion is to account for the response of the fluid to shearing, i.e. change in the flow's pattern in directions perpendicular to velocity. Such gradients arise in setups modelling one-dimensional expansion of matter, such as the studies discussed in Secs.~\ref{sec.lec2} and~\ref{sec.lec3}, that is why we elaborate on associated physics. More generally, the role of the shear viscosity contribution to the energy-momentum tensor, $- \eta(T) \, \sigma^{\mu \nu}$, is to homogenize the alignment of flow velocities. The primary role of $\eta$ in the experiment and HIC phenomenology lies in affecting the transversal (to the collision axis) expansion of plasma. This feature is absent in all the setups discussed in these lecture notes. See Refs.~\cite{vanderSchee:2012qj,vanderSchee:2013pia,Chesler:2015wra,Chesler:2016ceu} for holographic studies of transversal expansion in HICs.
}
\be
\label{eq.hydro1}
\Pi^{\mu \nu} = - \eta(T) \, \sigma^{\mu \nu} \, - \, \zeta(T) \, \Delta^{\mu \nu} \, \nabla_{\alpha} u^{\alpha},
\ee
contains a scalar function (transport coefficient), the shear viscosity $\eta(T)$, now known to have at least about an order of magnitude different value of the ratio with the equilibrium entropy density $s$ for weakly- and strongly-coupled~\cite{Policastro:2001yc} non-Abelian plasmas\footnote{Note that this is not the case with the thermodynamic properties captured by ${\cal E}(T)$ and $P(T)$, as has been known in holography from very early on. The weak-coupling value of $\eta/s$ quoted in Eq.~\eqref{eq.etas} is for ${\cal N} = 4$ SYM~\cite{Huot:2006ys} and is significantly smaller than for the pure Yang-Mills theory.}
\be
\label{eq.etas}
\frac{\eta}{s} \Big|_{\, \mathrm{small} \, \lambda} \approx \frac{6.174}{\lambda^{2} \log{\left( 2.36 / \sqrt{\lambda} \right)}} \quad \mathrm{vs.} \quad
\frac{\eta}{s} \Big|_{\, \lambda = \infty} = \frac{1}{4 \pi} \approx 0.08
\ee 
with $\lambda = g_{YM}^{2} N_{c}$ and the 't Hooft planar limit is implicitly implied throughout the text (we will refer to it as to the holographic regime). The successful description of the experimental signal requires including the shear viscosity with $\eta/s$ much closer to the famous strong-coupling result than to the leading order weak-coupling value, see, e.g., Ref.~\cite{Arnold:2007pg} for a discussion. This observation has provided the key impetus for the paradigm shift concluding that experimentally accessible QGP is a strongly-interacting liquid rather than a gas of quarks and gluons. Note also that, formally, relativistic hydrodynamics is phrased in the language of a systematic expansion in derivatives and Eq.~\ref{eq.hydro1} contain just the first subleading order. For a contemporary exposition of relativistic hydrodynamics, see Ref.~\cite{Baier:2007ix}.

The aim of the research program overviewed in these lectures and spanned by a series of work included as Refs.~\cite{Heller:2011ju,Heller:2012je,Heller:2012km,Heller:2013oxa,Casalderrey-Solana:2013aba,Casalderrey-Solana:2013sxa,Buchel:2015saa,Heller:2013fn,Heller:2015dha,Buchel:2016cbj} is to use holography to understand hydrodynamization, i.e. the process as a result of which the one-point functions of the energy-momentum tensor in a non-trivial state can be accurately described using hydrodynamic constitutive relations such as Eq.~\eqref{eq.hydro1}. Hydrodynamization implies vast reduction in the information about the microscopic system: mathematically speaking, in the relevant spacetime region 10 independent components of $\langle T^{\mu \nu} \rangle$ can be then well-approximated by only $4$ functions: $T$ and $u^{\mu}$ ($u_{\mu} u^{\mu} = - 1$). Note that in none of the exposition above appears the word ``equilibration'' and indeed the most interesting lesson stemming from these developments is that, remarkably,
\begin{center}
hydrodynamization $\neq$ equilibration,\\
\end{center}
see Sec.~\ref{sec.lec3} for details. In particular, the word ``hydrodynamization'' has been coined only recently precisely with these holographic developments in mind and was later showed, see Refs.~\cite{Kurkela:2015qoa,Keegan:2015avk,Heller:2016rtz}, to occur also within weakly-coupled frameworks. The phenomenological relevance of hydrodynamizations starts as a partial justification for the use of hydrodynamic models in HIC under omnipresent there extreme conditions (large gradients and/or small QGP droplets). For a further set of thought-provoking ideas on this front, see very recent Ref.~\cite{Romatschke:2016hle}.

The last key introductory remark is that the reason to use holography is certainly not that we naturally expect the strong-coupling physics to dominate the pre-hydrodynamic phase in HIC experiments at RHIC and LHC. It is rather the scarceness of methods to solve gauge theories in a time-dependent settings in a fully ab initio manner, i.e. without making ad hoc assumptions likely to affect the outcome, that made us explore this avenue. For completeness, let us mention two more fundamentally rooted in QCD weakly-coupled approaches to time-dependent processes: the so-called effective kinetic theory~\cite{Arnold:2002zm,Kurkela:2015qoa} and classical statistical approximation (see Secs.~10-12 of Ref.~\cite{Gelis:2015gza} for an overview of the latter). The past several years have seen many fascinating developments motivated by the desire to understand the emergence of the hydrodynamic description of HICs using these frameworks, but they run outside the scope of these lecture notes. It should, perhaps, also be stressed that analogous methods to the ones employed in the material discussed in these lectures in the context of HICs can be also applied to certain condensed matter theory problems, see, e.g., Ref.~\cite{Bhaseen:2012gg,Adams:2012pj,Sonner:2014tca} for examples of such applications.

The rest of these notes contain the write-up of three lectures presented by the author in Zakopane (Poland) in May 2016 as part of the 56$^\mathrm{th}$ Cracow School for Theoretical Physics ``A Panorama of Holography''. The original transparencies contain some complementary information and, as of October 2016, are available at
\begin{center}
\href{http://th-www.if.uj.edu.pl/school/2016/speakers.html}{http://th-www.if.uj.edu.pl/school/2016/speakers.html}
\end{center}
Part of this material was also covered by the author in Lisbon (Portugal) in July 2014 at the Summer School on 
String Theory and Holography. The lectures in Zakopane focused on the following aspects:
\begin{itemize}
\item {\bf hydrodynamization timescales} in strongly-coupled gauge-theories discussed in Sec.~\ref{sec.lec1},
\item {\bf holographic collisions} covered in Sec.~\ref{sec.lec2},
\item {\bf hydrodynamization} exposed in Sec.~\ref{sec.lec3}.
\end{itemize}
The material from the lectures is supplemented with Sec.~\ref{sec.methods} which streamlines some of the methodology and, at the same time, provides interested readers with information where to obtain the necessary know-how to contribute to this exciting line of research in the future. Sec.~\ref{sec.summary} briefly summarizes the main lessons from the series~\cite{Heller:2011ju,Heller:2012je,Heller:2012km,Heller:2013oxa,Casalderrey-Solana:2013aba,Casalderrey-Solana:2013sxa,Buchel:2015saa,Heller:2013fn,Heller:2015dha,Buchel:2016cbj}. The notes conclude with an idiosyncratic outlook section.

\section{Methodology and literature of the subject \label{sec.methods}}

The basic technical idea behind the developments described in these lectures is to analyze solutions of Einstein's equations with negative cosmological constant in (1+4)-dimensions
\be
\label{eq.Einstein}
R_{a b} - \frac{1}{2} R \, g_{a b} - \frac{6}{L^{2}} \, g_{a b} = 0
\ee
in terms of non-equilibrium processes occurring in dual (1+3)-dimensional gauge theories and imprinted in the form of the corresponding $\langle T^{\mu \nu} \rangle$. Eqs.~\eqref{eq.Einstein} imply studying conformal field theories (CFTs) with large number of microscopic constituents and strong interactions among them and states characterized only by $\langle T^{\mu \nu}\rangle$ with all other one-point functions vanishing. For definiteness, the example we will be always having in mind is ${\cal N} = 4$ SYM in the holographic regime, but it needs to be stressed that Eqs.~\eqref{eq.Einstein} actually describe a sector of infinitely many holographic CFTs~\cite{Bhattacharyya:2008mz}.

Both the conformal symmetry and the absence of other operators acquiring an expectation value can be relaxed using standard holographic technology, but since in the situations of interest (i.e. mimicking the dynamics of HICs) solving Eqs.~\eqref{eq.Einstein} turns out not to be completely straightforward we will not do so except from just quoting the results of Ref.~\cite{Buchel:2015saa}. In fact, understanding the implications of relaxing the conformal symmetry (absent in the QCD itself) on the studies discussed in these lectures is one of the most interesting open problems and a subject of many current investigations including Ref.~\cite{Buchel:2015saa}.
The general Ansatz for a solution of Eqs.~\eqref{eq.Einstein} can be written as
\be
\label{eq.AdS}
ds^{2} = \frac{L^{2}}{z^{2}} \left( dz^{2} + g_{\mu \nu} (z,x) \, dx^{\mu} dx^{\nu} \right).
\ee
The CFT vacuum in Minkowski space corresponds to $g_{\mu \nu}(z,x) = \eta_{\mu \nu}$ and excited states introduce non-trivial $z$- and $x^{\alpha}$-dependence in $g_{\mu \nu}$ but asymptote as $z \rightarrow 0$ to Minkowski metric $\eta_{\mu \nu}$ according to~\cite{deHaro:2000vlm}
\be
\label{eq.FG}
g_{\mu \nu}(z,x) = \eta_{\mu \nu} + \frac{2 \pi^{2}}{N_{c}^{2}} \, \langle T_{\mu \nu} \rangle (x) \, z^{4} + O(z^6).
\ee
In the equation above all the higher order terms are expressed in terms of $\langle T_{\mu \nu} \rangle$ and its derivatives containing derivatives of an arbitrary high order. Note that the above form of the metric justifies calling $x^{\alpha}$ the dual field theory coordinates in the sense that at the boundary these are the coordinates on the manifold (Minkowski space) in which the dual CFT lives. Note also that from the point of view of time evolution in AdS, we think about $\eta_{\mu \nu}$ as a boundary condition we fix (almost) at will, whereas $\langle T_{\mu \nu} \rangle$ is associated with the state and so is not completely arbitrary. One more remark is that the setup we want to study in a dual CFT is imprinted in our Ansatz on the form of $\langle T_{\mu \nu} \rangle$ and, through Eq.~\eqref{eq.FG}, this leads to the Ansatz for the corresponding bulk metric in coordinates~\eqref{eq.AdS}~\cite{Janik:2005zt}. Finally, let us emphasize that according to Eq.~\eqref{eq.FG} the constant $x^{0}$-slice in the bulk contains information about time derivatives of $\langle T_{\mu \nu} \rangle$ of an arbitrary order which means that at least formally the equations of motion for $\langle T_{\mu \nu} \rangle$ are of an infinite order~\cite{Beuf:2009cx}. Holography provides then a way to repackage these intractable relations into manageable second order partial differential equations at a price of introducing additional direction into the problem's description.

How to then answer the question we posed ourselves in the introduction? Conceptually, the situation is rather simple. When it comes to holography, we need to construct time-dependent solutions of Einstein's equations~\eqref{eq.Einstein}, bring them if needed to the form~\eqref{eq.AdS} known also as the Fefferman-Graham coordinates and, using series expansion, obtain the corresponding profiles of $\langle T^{\mu \nu} \rangle$ as functions of the $x^{\alpha}$-coordinates. The remaining task is a non-holographic one and lies in analyzing the form of $\langle T^{\mu \nu} \rangle$, in particular understanding how long it takes for it to be close to hydrodynamic predictions with no derivative corrections included~\eqref{eq.hydro0}, with only first derivative terms~\eqref{eq.hydro1}, etc. This will be the subject of Secs.~\ref{sec.lec1}~to~\ref{sec.lec3}. The rest of the current section discusses the former issue, i.e. obtaining the relevant (1+4)-dimensional geometries.

The gravity solution corresponding to global equilibrium\footnote{For a holographic gauge theory on a sphere (say $R\times S^{3}$) the situation is more complicated, see Ref.~\cite{Witten:1998zw}, but we will not be concerned with it since we study gauge theories on $R^{1,3}$.}, i.e., $\langle T^{\mu \nu} \rangle$ given by Eqs.~\eqref{eq.TmunuMAIN} and \eqref{eq.hydro0} with $T$ and $u^{\alpha}$ being constant and ${\cal E}(T) = \frac{3}{8} N_{c}^{2} \pi^{2} T^4$ and ${\cal P}(T) = \frac{1}{3} {\cal E}(T)$ is the AdS-Schwarzschild black brane~\cite{Witten:1998zw}
\be
\label{eq.bh}
ds^{2} = \frac{L^{2}}{z^{2}} \left\{ 
2 \, u_{\alpha} \, dx^{\alpha} dz - \left(1 - \pi^{4 }\, T^{4}  \, z^{4} \right) u_{\mu} u_{\nu} \, dx^{\mu} dx^{\nu} + \eta_{\mu \nu} \, dx^{\mu} dx^{\nu} 
\right\}.
\ee
For the sake of completeness of our discussion in this section, we wrote this solution in the ingoing Eddington-Finkelstein\footnote{Note that although, for simplicity, we denote the Eddington-Finkelstein AdS radial coordinate in, e.g., Eq.~\eqref{eq.bh} by $z$, it can be a priory a \emph{different} radial coordinate in AdS than the one appearing in the Fefferman-Graham Ansatz~\eqref{eq.AdS}. The same obviously applies to any other coordinate designations.} coordinates and interested reader can perform the relevant coordinate change to the Fefferman-Graham gauge and verify the corresponding form of~$\langle T^{\mu \nu} \rangle$. The black brane temperature is the temperature of the plasma phase in a dual CFT and the match extends to all the thermodynamic properties. The key feature of the AdS-Schwarzschild black brane is clearly the presence of the horizon at $r = \pi \, T \, L^{2}$ which plays a role of a perfectly absorbing membrane and is responsible for dissipation. 

This notion generalizes also outside global equilibrium and the processes of hydrodynamization in CFT is holographically related to, possible, horizon formation and its subsequent evolution. More precisely, the geometry of the AdS-Schwarzschild upon promoting $T$ and $u^{\alpha}$ to be functions of $x^{\beta}$ becomes the geometry dual to solutions of the perfect fluid hydrodynamics~\eqref{eq.hydro0}. Using this so-called fluid-gravity duality~\cite{Bhattacharyya:2008jc} one can work out subsequent corrections in derivatives of $T$ and $u^{\alpha}$ at first order matching viscous hydrodynamics (with in this case vanishing $\zeta$). 

\begin{figure}
\begin{center}
\includegraphics[width=9cm]{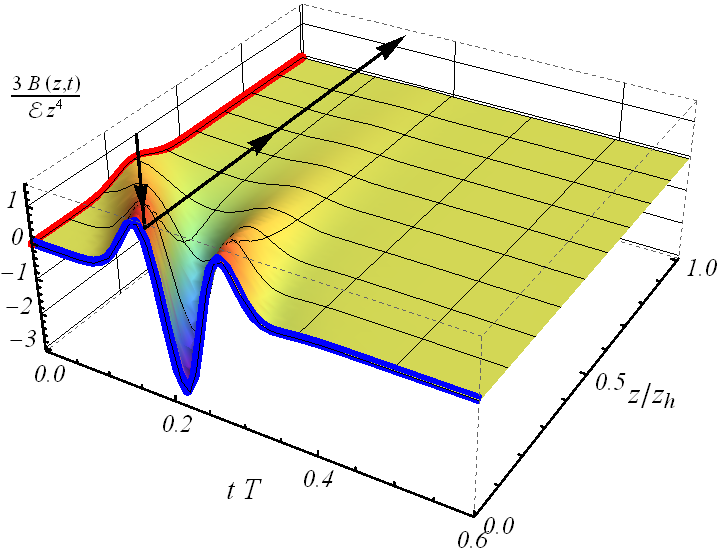}
\end{center}
\caption{The plot is taken from Ref.~\cite{Heller:2013oxa} (see also~Ref.~\cite{Heller:2012km}) and illustrates gravitational dynamics underlying holographic hydrodynamization in the homogeneous isotropization setup, see also Eq.~\eqref{eq.TmunuHomIso}, in which the bulk metric takes the form $ds^{2} = - \frac{2}{z^{2}} dt \, dz - A \,  dt^2 + \Sigma^2 e^{- 2 B} \, dx_{1}^{2} + \Sigma^2 e^{B} \,  \left( dx_{2}^{2} + dx_{3}^{2} \right)$ and $A$, $\Sigma$ and $B$ are functions of the ingoing Eddington-Finkelstein time $t$ and the AdS radial coordinate~$z$. Note at this point that the adopted metric Ansatz follows from Eqs.~\eqref{eq.AdS} and~\eqref{eq.FG} given the form of $\langle T^{\mu \nu} \rangle$ in Eq.~\eqref{eq.TmunuHomIso}. The plot depicts the function $B$ directly capturing pressure anisotropy in a dual CFT normalized by its leading near-boundary ($z=0$) behaviour and the equilibrium pressure, ${\cal P} = \frac{1}{3} {\cal E}$, as a function of the Eddington-Finkelstein time $t$ and the radial coordinate $z$. Note that slices of constant $t$ are null. The red curve depicts Gaussian initial data. The Gaussian splits into an outgoing (black arrow) and ingoing wave packets. The outgoing wave packet hits the boundary at $z = 0$, gets reflected and is immediately (in the coordinate time $t$, see the arrow) absorbed by the horizon residing in the vicinity of $z = z_{h}$. The same happens to the initial ingoing wave packet, i.e. it is absorbed by the horizon already at $t = 0$. The blue curve denotes the pressure anisotropy in the dual CFT normalized to the equilibrium pressure, $\frac{\Delta {\cal P}(t)}{{\cal E}/3}$, as a function of time $t$. One clearly sees hydrodynamization already at $t \approx 0.35/T$. Finally, note that despite apparent initial isotropy, the system is certainly far from equilibrium and this can be seen in a CFT by examining nonlocal correlation functions, such as $\langle \, T_{11}(t,\vec{x}\,) \, T_{11}(t,\vec{x} + \vec{d}\,) \rangle$ with nonzero $\vec{d}$.
\label{fig.holoiso}}
\end{figure}

We can precisely define the holographic dual to hydrodynamization as a phenomenon in which a gravity solution is going to be locally, in the sense of tubes spanned by constant values of $x^{\alpha}$ Eddington-Finkelstein coordinates, described by the geometry of the fluid-gravity duality. The picture with the aforementioned tubes can be advocated using bulk causality, see Ref.~\cite{Bhattacharyya:2008mz}. Fig.~\ref{fig.holoiso} provides some intuition on the relevant gravitational dynamics using the example of homogeneous isotropization discussed in the next section.

%The left figure shows $B$ as a function of time and radial coordinate for the initial profile $B_{ini}(z)=\frac{a_{4}}{10}z^{4}e^{-150(z-a_{4}^{1/4}/3)^{2}}$ which is initially localized near $z=a_{4}^{1/4}/3$. The right figure shows the Kretschmann scalar (subtracted by its value in the final state, i.e. for each $z$ we subtract corresponding value of the Kretschmann scalar for the static AdS Schwarzschild) as a function of time and radial coordinate for the same initial profile. One clearly sees on this plot the wave bouncing off the boundary and falling into the black brane. In the adopted generalized ingoing Eddington-Finkelstein coordinates this happens instantaneously. The blue curve imprinted on the boundary depicts the pressure anisotropy as a function of time in a dual field theory, whereas magenta curve approximates it using linearized Einstein's equations.

Of course, we are not interested in understanding these particular geometries since they are already known once we feed them with particular solutions of hydrodynamics. What we want to do is to generate geometries that at least initially are \emph{not} described by the fluid-gravity duality and there are 3 distinct ways to obtain them:
\begin{enumerate}
\item {\bf Starting with the vacuum and turning on CFT sources to excite it.} In the context of pure gravity in AdS studied here, this amounts to putting the dual CFT in a nontrivial background spacetime. In practical terms, one chooses the leading small-$z$ behaviour of $g_{\mu \nu}(z,x)$ from Eq.~\eqref{eq.AdS} accordingly. Note that this is how the numerical holography started, see the pioneering Refs.~\cite{Chesler:2008hg,Chesler:2009cy}, but we will not pursue this line here since it turns out to be tricky to disseminate between the process of creating an excited state and the subsequent description of $\langle T^{\mu \nu} \rangle$ using the laws of hydrodynamics.
\item {\bf Superposing separated localized exact solutions of Einstein's equations.} An example of such solutions directly relevant for the material covered in Sec.~\ref{sec.lec2} are gravitational shock-waves
\be
\label{eq.shocks}
ds^{2} = \frac{L^{2}}{z^{2}} \left( dz^{2} + \eta_{\mu \nu} dx^{\mu} dx^{\nu} + z^{4} h_{\pm}(x_{\pm}) dx_{\pm}^{2} \right),
\ee
where $h_{\pm}$ is an arbitrary function $\geq 0$ and $x_{\pm} = x^{0} \pm x^{1}$~\cite{Janik:2005zt,Chesler:2010bi}. Each such a shock-wave (i.e. a left- or a right-moving excitation) is an exact solution of the AdS gravity, but a linear combination of approaching shocks ceases to solve the Einstein's equations in AdS once nontrivial overlaps in the bulk start developing. The advantage of this approach is that it allows to study collisions, the obvious drawback is that the class of the simplest projectiles \eqref{eq.shocks} might be too restrictive in the context of HIC phenomenology, see in particular Ref.~\cite{vanderSchee:2015rta}. Also, shocks~\eqref{eq.shocks} move precisely at the speed of light, whereas in HICs the projectiles are boosted nuclei with large but finite $\gamma$-factor.
\item {\bf Solving initial value problem in AdS for arbitrary initial metric.} In this case, one \emph{defines} excited states in terms of the initial data specified on some constant-time surface in the bulk of AdS. The initial data have to satisfy the relevant constraint equations and lead to spacetimes free of naked singularities. Since the CFT metric remains flat at all times, there is a clear separation between hydrodynamic and non-hydrodynamic parts of the evolution. The main disadvantage is that one looses clear CFT interpretation of considered non-equilibrium states as opposed to the first approach. The method of choice to minimize this drawback was to scan a large number of different states and identify universal features of the dynamics.
\end{enumerate}

Lecture 1 in Sec.~\ref{sec.lec1} will implicitly use the first method in the context of the linear response theory. Lecture 2 in Sec.~\ref{sec.lec2} will focus exclusively on the results delivered using the second approach. Most of the presented results in lectures~1 and~3 (Sec.~\ref{sec.lec3}) will adopt the third approach and scan many different states.

Vast majority of the developments discussed in these lectures are based on numerical relativity methods, i.e. solving Einstein's equations numerically on computers. Sometimes such developments in the context of AdS gravity are referred to as to ``numerical holography''. The common technical denominator of the problems considered in these lectures that makes them very special is the horizon spanning all the $x^{\mu}$-directions. In such cases, the method of choice is to use generalized ingoing Eddington-Finkelstein coordinates and pseudospectral discretization. The former one leads to a very stable numerics since all the signals moving inwards in AdS get immediately (in the Eddington-Finkelstein ``time'' in which one evolves) absorbed by the horizon, whereas the latter ensures very good accuracy at modest grid sizes resulting in simulations of spacetimes depending on all $1+4$ coordinates and at the same time being within the range of powerful yet still desktop computers~\cite{Chesler:2015wra}. A very accessible introduction to these two methods combined can be found in the excellent overview article~\cite{Chesler:2013lia} by Paul Chesler and Larry Yaffe who pioneered their use in the context of holography. Numerous aspects of pseudospectral methods when used in numerical relativity are also discussed in, e.g., Ref.~\cite{Grandclement:2007sb}. Another useful resource on this front is the Ph.D. thesis of my collaborator on these topics, Wilke van der Schee, available as Ref.~\cite{vanderSchee:2014qwa} as well as materials on his website including runnable Mathematica programs simulating black brane evolution in AdS space\footnote{As of October 2016 available at: \href{https://sites.google.com/site/wilkevanderschee/ads-numerics}{https://sites.google.com/site/wilkevanderschee/ads-numerics}}. Note that some of the most interesting open problems fail to fall in the class amenable to such treatment and for them more standard and resource-consuming approaches are needed. An example of such a problem is a collision of two black holes in AdS space considered in Ref.~\cite{Bantilan:2014sra}. Finally, some readers might find Ref.~\cite{Heller:2012je} useful which discusses and utilizes a more standard approach to the initial value problem in general relativity in the context of perhaps the simplest hydrodynamization process being Bjorken's boost-invariant flow \cite{Bjorken:1982qr} (see Sec.~\ref{sec.lec3} for more information about the latter). Similar studies were performed later on also in the Eddington-Finkelstein coordinates in Ref.~\cite{Jankowski:2014lna}. 

Finally, let me mention several very good reviews of applications of holography to non-equilibrium processes in gauge theories. Interested reader is invited to consult Refs.~\cite{CasalderreySolana:2011us,DeWolfe:2013cua,Chesler:2015lsa}. 

\section{Lecture 1: hydrodynamization timescales at strong coupling \label{sec.lec1}}

The first question of an obvious interest to ask is how long it takes for a strongly-coupled gauge theory to reach hydrodynamic regime as seen by its $\langle T^{\mu \nu} \rangle$ starting from some general non-equilibrium state. The simplest setting to answer this was the linear response theory, i.e. the leading order response of the thermal state to external agents encapsulated by quantum field theory sources. In the case of $T^{\mu \nu}$, the corresponding source is the background metric $\eta_{\mu \nu} + \delta g_{\mu \nu}(x)$ and the relevant formula is the following
\be
\delta \langle T^{\mu \nu} \rangle = \int d^{3} k \int_{-\infty}^{\infty} d\omega \, e^{-i \, \omega \, t + i \,  \vec{k} \cdot \vec{x}} \, \left[ \mathrm{res}_{\omega = \omega_{mode} (k)}G_{R}(\omega, \vec{k}) \cdot \delta g(\omega_{mode}, \vec{k}) \right]\,^{\mu \nu}.
\ee
In the equation above we work in the Fourier space, $G_{R}$ is the retarded two-point function of the energy-momentum tensor in the thermal state and $\left[G_{R}(\omega, \vec{k}) \cdot \delta g(\omega, \vec{k})\right]\,^{\mu \nu}$ encapsulates a particular contraction pattern\footnote{Note that for every value of the momentum $\vec{k}$, $G_{R}$ splits into three independent channels transforming irreducibly under rotations with respect to the axis set by $\vec{k}$. We will ignore this important feature in order not to make the presentation too technical. Also, the discussion in this section misses the important question of the actual form of residues of $G_{R}(\omega, \vec{k})$ and interested reader should consult in this respect Ref.~\cite{Amado:2007yr}.}. Using the standard complex analysis methods one can rephrase the integral over the frequency in terms of a sum over singularities of $G_{R}(\omega, \vec{k})$ on the complex plane. A highly nontrivial prediction of the Einstein gravity holography from the quantum field theory point of view is that these singularities have a form of an infinite number of single poles for each value of~$\vec{k}$:
\be
\label{eq.TmunuMODES}
\hspace{-10 pt}\delta \langle T^{\mu \nu} \rangle = \int d^{3} k \sum_{modes} e^{-i \, \omega_{mode}(k) \, t + i \,  \vec{k} \cdot \vec{x}} \,  \left[ \mathrm{res}_{\omega = \omega_{mode} (k)} G_{R} (\omega, \, \vec{k}) \cdot \delta g(\omega_{mode}(k), \vec{k}) \right]\,^{\mu \nu}.
\ee
The frequencies $\omega_{mode}(k)$ all lie in the lower-half plane of the complex-$\omega$ plane, which leads to the decay of those modes: $e^{- | \mathrm{Im} \, \omega_{mode}(k) | \, t}$. Gravitational interpretation of all these excitations is that of quasinormal modes (QNMs) of the AdS-Schwarzschild black brane~\cite{Kovtun:2005ev}, see Ref.~\cite{Berti:2009kk} for a comprehensive review of QNMs in black hole physics including holography.

All the QNMs apart from two are gapped, i.e. $\omega_{mode}(k = 0) \neq 0$ with a nontrivial imaginary part. The remaining two modes can be made arbitrarily long-lived by adding support for $\delta g(\omega, \vec{k})$ at very small $k$. They correspond to linearized hydrodynamic excitations, shear and sound waves, and their gapless nature can be traced back to their equations of motion being sorely conservation equation for $\langle T^{\mu \nu} \rangle$ satisfying hydrodynamic constitutive relations like Eq.~\eqref{eq.hydro1} and its higher order generalizations. In the rest of these notes, we will refer to transient modes as to the massive QNMs and to the long-lived modes as to the hydrodynamic modes.

\begin{figure}
\begin{centering}
\includegraphics[width=7cm]{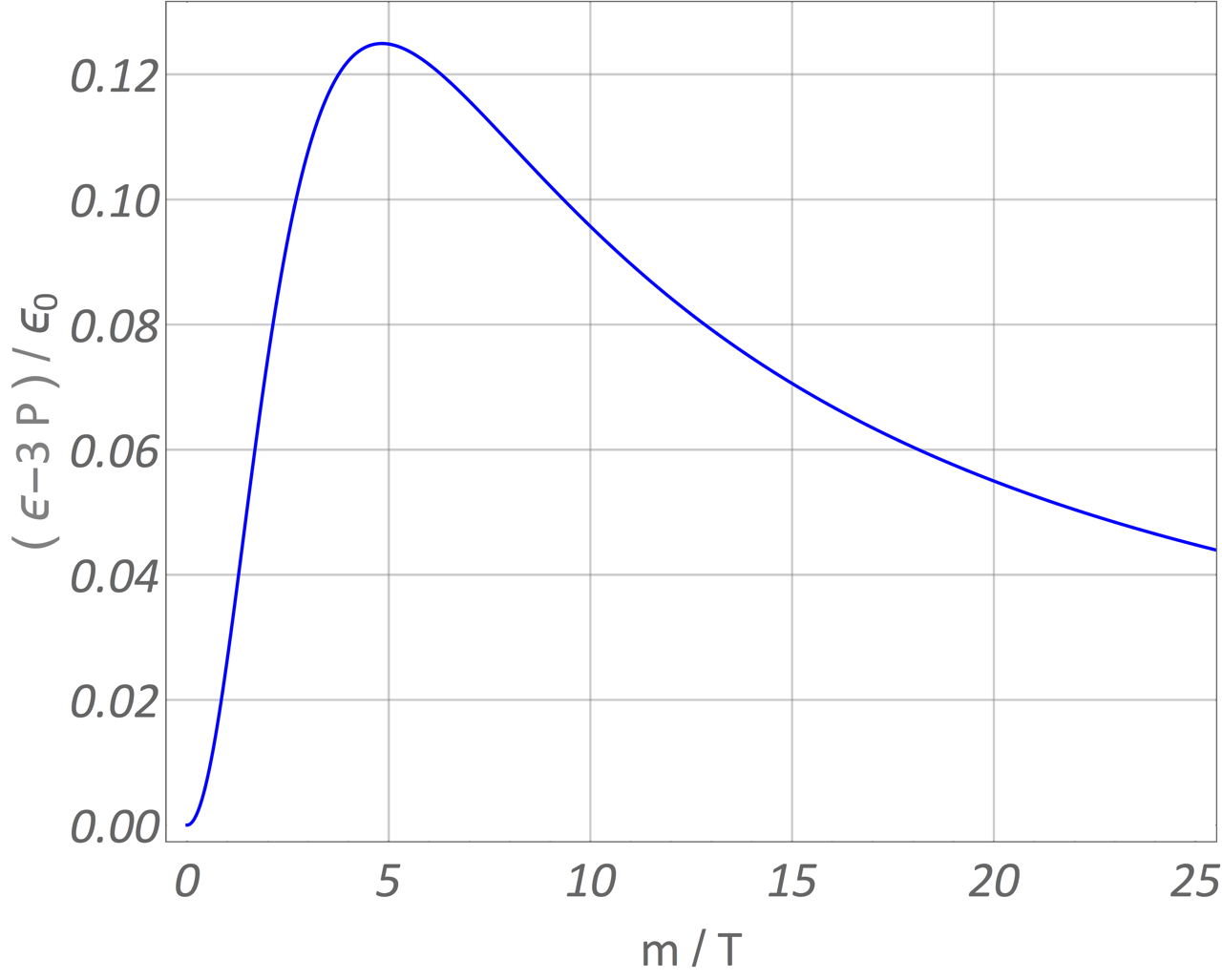}\quad \includegraphics[width=7cm]{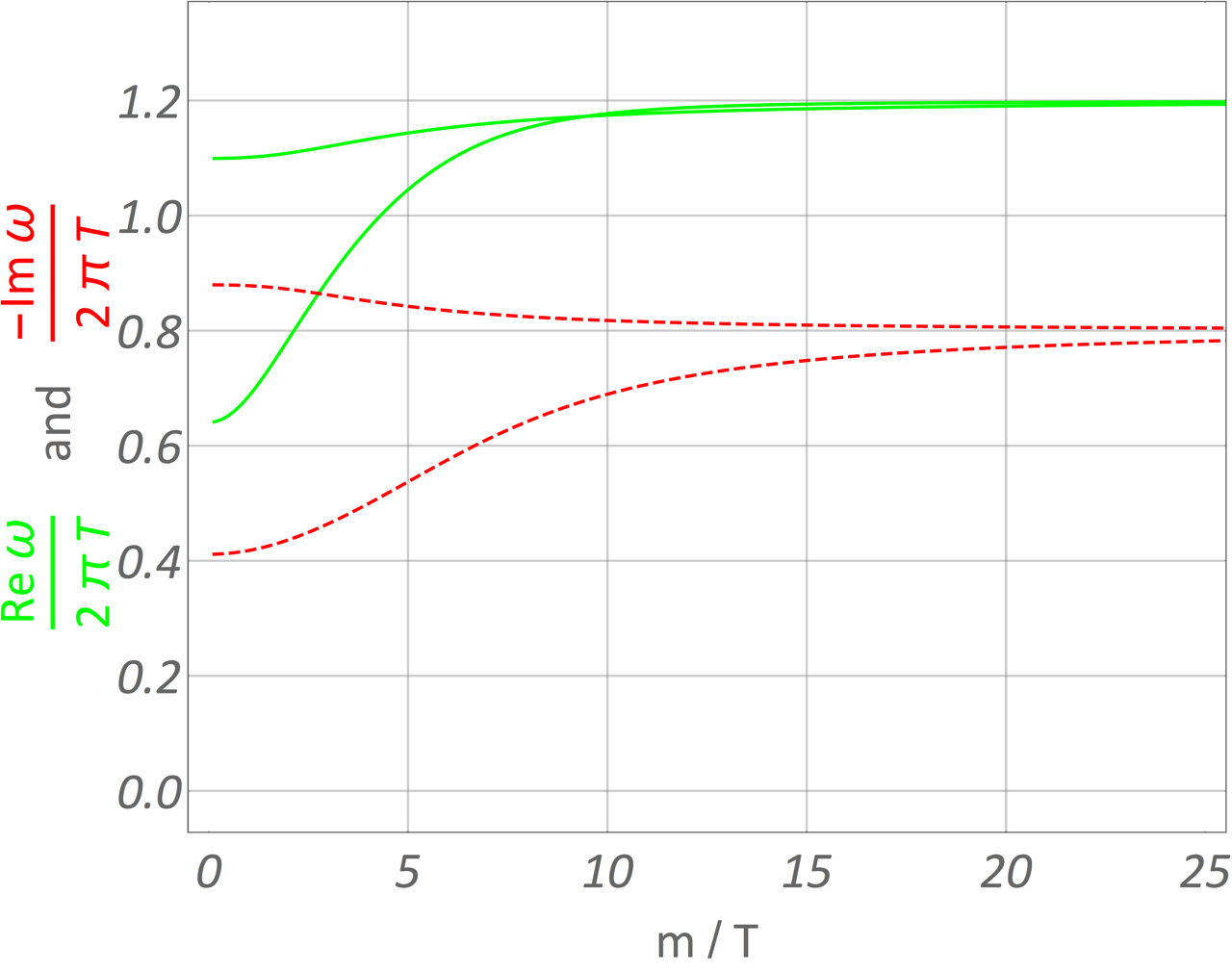}
\par\end{centering}
\caption{\label{fig.N2S} The results taken from Ref.~\cite{Buchel:2015saa} illustrate the thermodynamics and non-equilibrium properties of ${\cal N} = 2^{*}$ gauge theory. {\bf Left:} trace of the equilibrium energy-momentum tensor normalized to ${\cal N} =4$ SYM energy density at the same temperature denoted ${\cal E}_{0}$ as a function of the conformal symmetry breaking parameter $m$ normalized to the temperature $T$. According to this measure, the biggest deviations from the conformal invariance occur for $m/T \approx 4.8$. {\bf Right:} the frequencies of the least damped QNMs of $\Delta = 3$ and $\Delta = 2$ scalar operators (from top to bottom) present in ${\cal N} = 2^*$ gauge theory evaluated at vanishing momentum and expressed as functions of $m/T$. The key thing to notice is that the frequencies of these modes, as well as all the other ones considered in the original reference, do not change \emph{significantly} as a function of $m/T$. As a result, one does not expect parametric changes in hydrodynamization rates. This finding was corroborated by a later study in Ref.~\cite{Buchel:2015ofa} which considered a holographic setup whose thermodynamics is even more affected by the conformal symmetry breaking, as well as Refs. mentioned in the main body of the text.}
\end{figure}

Note now that the vacuum Einstein's equations \eqref{eq.Einstein} do not contain any non-trivial dimensionless parameters, nor does the AdS-Schwarzschild black brane solution \eqref{eq.bh}. The latter is fully specified by its temperature $T$ and it is then rather natural to expect that massive QNMs have $\mathrm{Im} \, \omega(k)$ of the order of the temperature. This means that unless the occupation number in Eq.~\eqref{eq.TmunuMODES}, i.e. the term $\left[ \mathrm{res}_{\omega = \omega_{mode} (k)} G_{R} (\omega, \, \vec{k}) \cdot \delta g(\omega_{mode}(k), \vec{k}) \right]\,^{\mu \nu}$, is parametrically enhanced all the massive QNMs will become negligible after the time scale set by the temperature. Since this possibility, at least superficially, is ruled out by the virtue of considering small perturbations, one trivially arrives at the conclusion that for small-in-amplitude perturbations of the equilibrium plasma hydrodynamic description becomes the only relevant one after timescale set by the temperature inverse:
\be
\label{eq.thydroT}
t_{hydro} \, T = O(1).
\ee
The rest of this lecture is devoted to elucidating the influence on the above relation of introducing additional scales to the problem via:
\begin{itemize}
\item Breaking the conformal symmetry but considering small perturbations of the equilibrium plasma~\cite{Buchel:2015saa};
\item Keeping the conformal symmetry intact but looking at large amplitude perturbations in a simple-to-interpret setup~\cite{Heller:2012km,Heller:2013oxa}.
\end{itemize}
Regarding breaking the conformal symmetry, the idea is to start with a holographic CFT and turn on sources for some relevant operator(s). We will look at a particular way of breaking the conformal symmetry, which preserves some of the supersymmetry of ${\cal N} = 4$ SYM. As a result of using a top-down model\footnote{Gravitational theory being dual to a at least a sector of a known quantum field theory.}, the so-called ${\cal N} = 2^{*}$ gauge theory~\cite{Pilch:2000ue,Buchel:2003ah,Buchel:2007vy}, we do not have freedom in introducing ranges of dimensionless parameters to the gravity-coupled-to-matter equations of motion. Skipping the details, the take home message from the analysis originally present in Ref.~\cite{Buchel:2015saa} is that the relation~\eqref{eq.thydroT} holds at the level of small perturbations of equilibrium plasma also in ${\cal N} = 2^{*}$ gauge theory, see Fig.~\ref{fig.N2S}. Combining this with the results of other researchers, in particular Refs.~\cite{Janik:2015waa,Fuini:2015hba,Ishii:2015gia,Attems:2016ugt,Attems:2016tby}, points towards the conclusion that Eq.~\eqref{eq.thydroT} might hold more generally when the conformal symmetry is broken. Finally, note that it does not imply that breaking the conformal symmetry does not lead to new interesting physics effects absent in holographic CFTs, see, e.g., Ref.~\cite{Attems:2016tby}.

Regarding adding new scales through initial conditions, it turns out that the simplest setup to achieve this is the so-called homogeneous isotropization~\cite{Chesler:2008hg} in which the $\langle T^{\mu \nu} \rangle$ takes the form
\be
\label{eq.TmunuHomIso}
\langle T^{\mu \nu} \rangle = \mathrm{diag} \left\{ {\cal E}, \, \frac{1}{3}{\cal E} - \frac{2}{3} \Delta {\cal P}(t), \, \frac{1}{3}{\cal E} + \frac{1}{3} \Delta {\cal P}(t), \, \frac{1}{3}{\cal E} + \frac{1}{3} \Delta {\cal P}(t) \, \right\}\, ^{\mu \nu}.
\ee
In this setup, the energy density is constant due to conservation of the energy-momentum tensor and the pressure anisotropy $\Delta {\cal P} (t)$ is the transient effect of interest. The simplest dimensionless quantity associated with the form of $\langle T^{\mu \nu} \rangle$ from Eq.~\eqref{eq.TmunuHomIso} is $\Delta {\cal P}(t) / {\cal E}$. Since $\Delta {\cal P}(t)$ vanishes in equilibrium (see also Ref.~\cite{Janik:2008tc}), we define the hydrodynamization time as the instance of the evolution after which the magnitude of the ratio $\Delta {\cal P}(t) / {\cal E}$ is lower than some threshold value, e.g.~10\%. Note also that the homogeneous isotropization does not excite hydrodynamic modes since there is not momentum in the system: in Eq.~\eqref{eq.TmunuMODES} this corresponds to setting $\vec{k} = 0$ and summing only over the massive QNMs. A very special property of this setup is that we know the final state from the start -- it is going to be the equilibrium plasma of the energy density we start with. 

The key idea behind the studies presented in Refs.~\cite{Heller:2012km,Heller:2013oxa} was to, first, make use of the simplicity of this setup to analyze very many different initial states in a comprehensive manner and, second, each time compare the results of the evolution using full nonlinear Einstein's equations with negative cosmological constant~\eqref{eq.Einstein} with the linear response theory. The reason for the latter comparison was to understand how important are nonlinear effects --  interactions between QNMs. Since due to reasons discussed in Sec.~\ref{sec.methods} we do not use the metric in which a CFT lives to perturb the system, what is meant here by the linear response theory are solutions of Einstein's equations linearized on top of the final state -- static AdS-Schwarzschild black brane with temperature set by the energy density -- fed with \emph{the same} initial conditions as the corresponding fully nonlinear evolution. This turns out to be a rare luxury since in the most interesting cases being expanding plasma systems discussed in Secs.~\ref{sec.lec2} and~\ref{sec.lec3} we do not seem to have a simple way to obtain the final state on top of which we could linearize Einstein's equations and understand the importance of nonlinear transient effects.

\begin{figure}
\begin{centering}
\includegraphics[width=9cm]{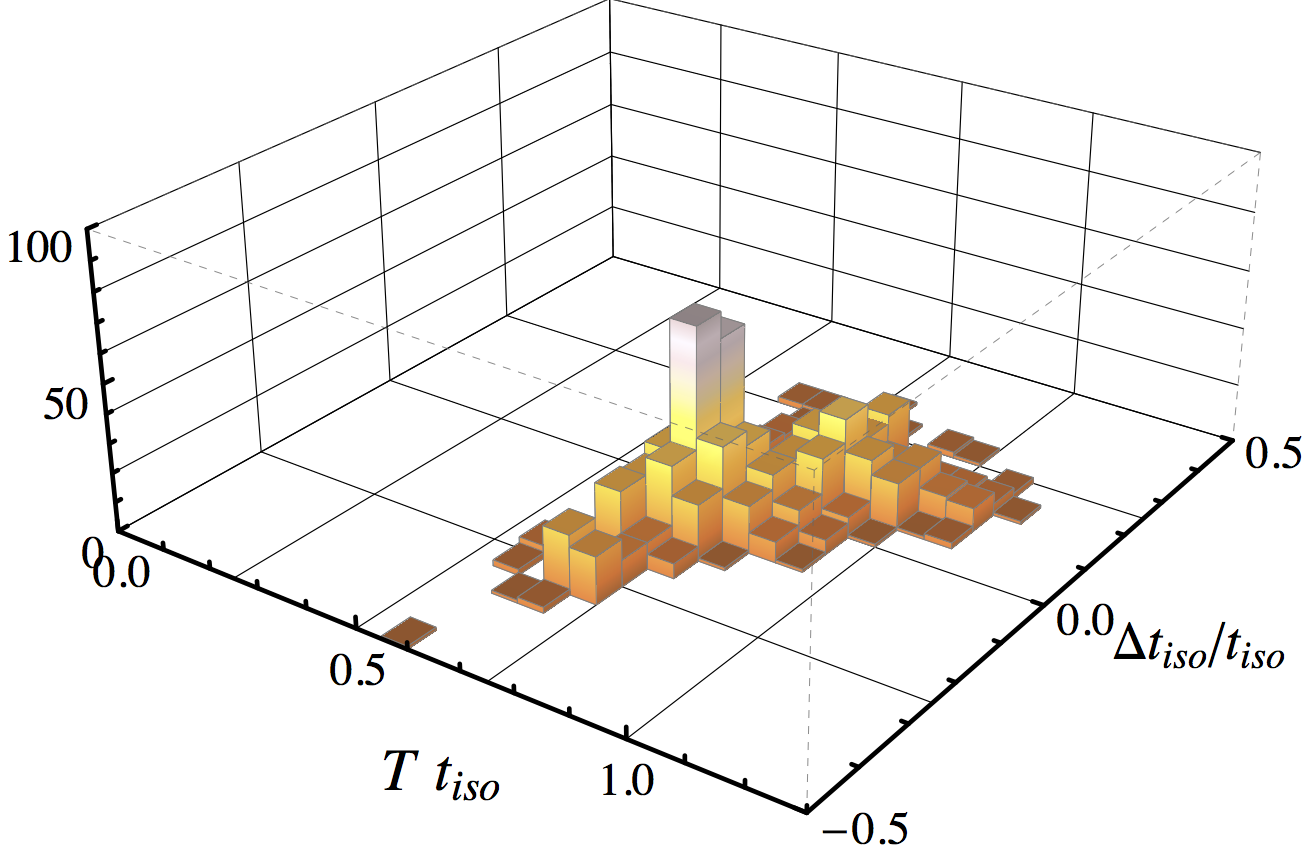}\\
\vspace{10 pt}
 \includegraphics[width=9cm]{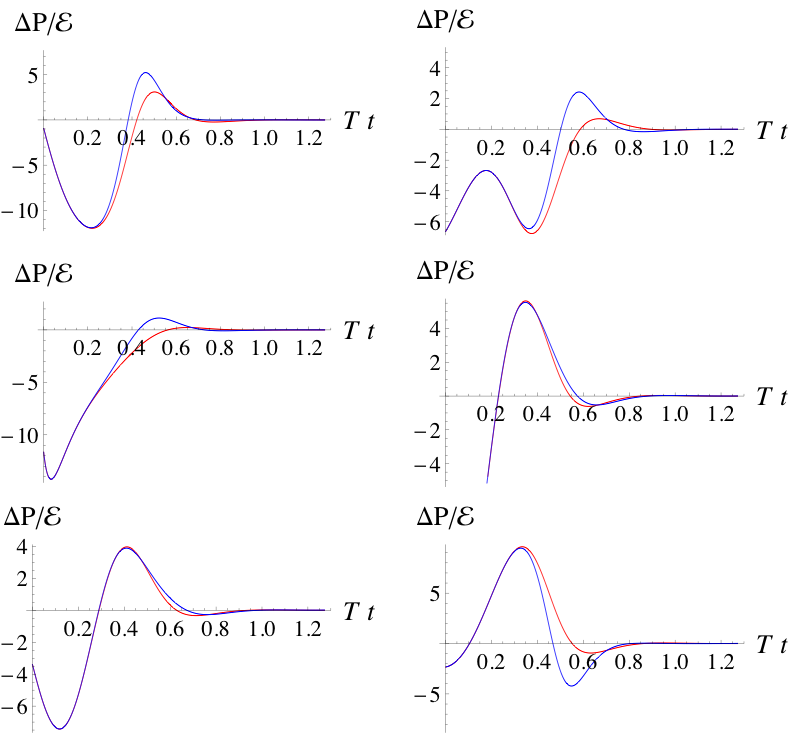}
\par\end{centering}
\caption{{\bf Top:} The histogram shows that none of the considered 800 different initial states takes longer than about $1.1/T$ to reach hydrodynamic regime in the setup of homogeneous isotropization. Furthermore, the plot also shows rather impressive agreement between the results of nonlinear Einstein's equations and the linear response theory -- the difference between predicted isotropization times, $\Delta t_{iso}$, is never large. {\bf Bottom:} Explicit comparison between pressure anisotropy as function of time obtained from solutions of nonlinear Einstein's equations (blue curves) and using linear response theory (red curves). One sees impressive semi-quantitative agreement even in the regime of very large anisotropies, i.e. outside the naive regime of validity of the linear response theory. All the plots in the figure are taken from Ref.~\cite{Heller:2013oxa}.
}
\label{fig.histogram}
\end{figure}

We still want to interpret the prediction of linearized equations in terms of a sum over QNMs, but now instead of the term $\left[ \mathrm{res}_{\omega = \omega_{mode} (k = 0)} G_{R} (\omega, \, \vec{k} = 0) \cdot \delta g(\omega_{mode}(k = 0), \vec{k} = 0) \right]\,^{\mu \nu}$ from Eq.~\eqref{eq.TmunuMODES} we just have a numerical coefficient that is encoded in the initial data in a rather intricate way, see Ref.~\cite{Heller:2013oxa}. The take-home message from these studies, see Fig.~\ref{fig.histogram}, is that despite initial conditions contain infinitely many scales (given by all the derivatives of $\Delta {\cal P}(t)$ at $t = 0$), the hydrodynamization in this setup\footnote{Comparing with Eq.~\eqref{eq.TmunuMAIN}, we immediately obtain that for homogeneous isotropization $T$ and $u^{\mu}$ are constant with $u^{\mu} \partial_{\mu} = \partial_{t}$. This implies that the gradient expansion is trivial to all orders, i.e. constant in whole spacetime and \emph{isotropic} perfect fluid $\langle T^{\mu \nu} \rangle$ is the exact prediction of hydrodynamics in this setup. This justifies \emph{in this particular case} using isotopization of pressures in $\langle T^{\mu \nu} \rangle$ as a criterium for the applicability of hydrodynamics.} still occurs over a timescale of 1/T and linear response theory used outside its naive regime of validity does a remarkable job in reproducing the features of $\langle T^{\mu \nu} \rangle$, see Fig.~\ref{fig.histogram}. It needs to be stressed that this apparent university of hydrodynamization time, confirmed in many other setups including those discussed in subsequent lectures, is not yet very well understood, see Ref.~\cite{Fuini:2015hba} for a nice toy-model. 

The lessons from this lecture are very interesting phenomenologically since Eq.~\eqref{eq.thydroT} is in agreement with the values used in hydrodynamic modelling of HICs at RHIC and LHC, where the role of $T$ is played by the local temperature at the moment of the applicability of hydrodynamics as noticed already in Ref.~\cite{Chesler:2008hg}. Although we did not excite nonlinear hydrodynamics in the setups discussed here, the results discussed in Secs.~\ref{sec.lec2} and~\ref{sec.lec3} indicate that Eq.~\eqref{eq.thydroT} remains valid also in these more realistic cases. Of course, all this does not imply that the approach to hydrodynamics in HICs is governed by the strong coupling processes, but rather that if it were, then the natural timescale for the applicability of hydrodynamics would be parametrically given by $1/T$ evaluated at the moment the hydrodynamics kicks~in. Furthermore, any serious phenomenological applications require better understanding the proportionality constant in Eq.~\eqref{eq.thydroT} which varies (even by a factor of~$2$) depending on strongly-coupled gauge theory, setup and initial state. Finally, let us note that the hydrodynamization timescale from Eq.~\eqref{eq.thydroT} is sometimes stated as a possible strong-coupling resolution of the so-called thermalization puzzle in HICs (see Ref.~\cite{Heinz:2004pj} for an early discussion of this point), i.e. the failure of extrapolated\footnote{Meaning leading order results evaluated at intermediate values of the coupling, outside the naive regime of validity.} weak-coupling calculations of \emph{thermalization} time to match the timescales needed to be used by nuclear theorists to start hydrodynamic modelling. As we will see in Sec.~\ref{sec.lec3}, there might not be a puzzle after all since (local) thermalization is a priori a different process than hydrodynamization and the thermalization timescale can be parametrically bigger than the time needed for the applicability of hydrodynamics

\section{Lecture 2: holographic collisions \label{sec.lec2}}

One of the most impressive outcomes of numerical holography are collisions of gravitational shock-waves~\eqref{eq.shocks} initiated in Ref.~\cite{Chesler:2010bi}. The latest developments on this front are universality in the late-time behaviour of plasma originating from planar shock-waves collisions~\cite{Chesler:2015fpa}, off-central collisions of objects with localized transversal structure~\cite{Chesler:2015wra}, collisions of planar objects carrying conserved charge mimicking the baryon number~\cite{Casalderrey-Solana:2016xfq} and collisions of planar shock-waves in strongly-coupled gauge theories with broken conformal symmetry~\cite{Attems:2016tby}. The aim of this lecture is to present those aspects of the planar shock-waves collisions from Eq.~\eqref{eq.shocks} discussed in Refs.~\cite{Casalderrey-Solana:2013aba,Casalderrey-Solana:2013sxa} that seem the most relevant in the light of newer developments in this area. 

\begin{figure}
\begin{center}
\includegraphics[width=0.6 \textwidth]{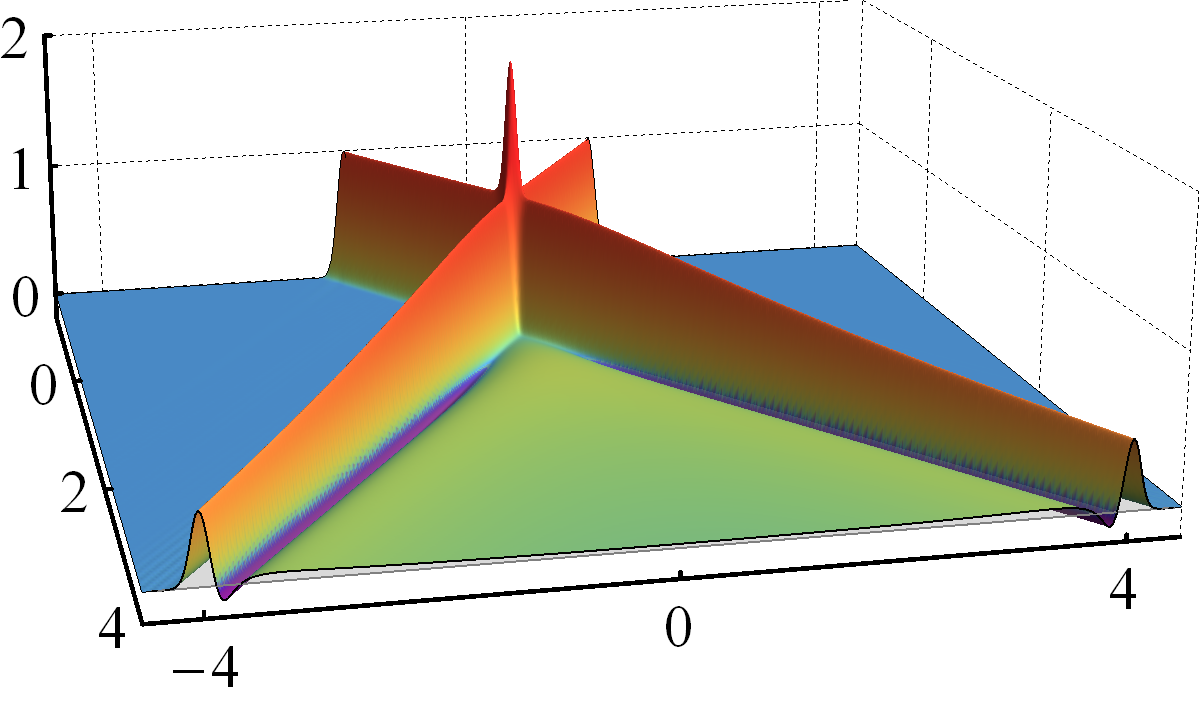}
\put(-167,160){\mbox{\large ${\cal E}/\rho^4$}}
\put(-275,45){\mbox{\large $\rho \, x^{0}$}}
\put(-120,-10){\mbox{\large $\rho \, x^{1}$}}
\end{center}
\caption{The lab-frame energy density, ${\cal E} \equiv \frac{2 \pi^{2}}{N_{c}^{2}} \langle T^{00}\rangle$, as a function of the lab-frame time $x^{0}$ and the longitudinal position $x^{1}$ for colliding shocks characterized by $h_{\pm} (x^{\pm}) = \rho^{4} \exp{\left[-\frac{1}{2\,w^{2}}\left( x^{\pm}\right)^2 \right]}$ and $\rho \, w \approx 0.08$. Note that there is no dynamics in the transversal plane. Regarding the actual dynamics, at the back of the plot one sees approaching maxima of the energy density. The collision point correspond with a very good accuracy to the overall maximum of the energy density. In the future light cone one sees two decaying remnants of the projectiles separated from the created matter by the transient region of negative energy density. At $x^{1} = 0$ hydrodynamics becomes a good description already at about $\rho \, x^{0} \approx 1.5$ which turns out to agree with Eq.~\eqref{eq.thydroT}, see Sec.~\ref{sec.lec3} for an extensive discussion of hydrodynamization in the context of expanding plasma systems. Finally, note that in the vicinity of the future light cone the system is so far from equilibrium that there exists no reference frame in which the matter if co-moving, i.e. $\langle T^{\mu \nu} \rangle$ does not possess there a time-like eigenvector. The plot is taken from Ref.~\cite{Casalderrey-Solana:2013aba}.}
\label{fig.shocks}
\end{figure}

Regardless of the choice of projectiles, i.e. the choice of the longitudinal structure of projectiles encapsulated by functions $h_{\pm}(x^{\pm})$ in Eq.~\eqref{eq.shocks}, collisions of shocks in Refs.~\cite{Casalderrey-Solana:2013aba,Casalderrey-Solana:2013sxa} always resulted in hydrodynamization with the projectiles either completely dissolved or decaying, see Fig.~\ref{fig.shocks}. Furthermore, Ref.~\cite{Chesler:2015fpa} was able to perform numerical simulations in all the cases of interest till the complete decay of projectiles remnants finding out that the resulting \emph{late-time} temperature and velocity distribution in the $x^{0}-x^{1}$ plane are universal\footnote{It is not entirely clear whether this decay of projectiles is a universal feature of holographic setups since all holographic collisions performed to date involve essentially the same type of projectiles. The general argument for it might be though the following: the horizon in general relativity cannot break so if the projectile remnants are attached to the horizon associated with what later becomes hydrodynamized plasma they are destined to decay.}. This finding encapsulates earlier claims made in Ref.~\cite{Casalderrey-Solana:2013sxa} and based on observed universalities in the hydrodynamized plasma for the collisions of sufficiently ``thin'' (at fixed total energy per unit transversal area or peak energy density) shock-waves. Although shock-waves collisions allow to make the statement that hydrodynamization $\neq$ thermalization, in Sec.~\ref{sec.lec3} we will use simpler and hence much cleaner example of the boost-invariant flow to discuss it.

As a result, the most interesting phenomena observed in Ref.~\cite{Casalderrey-Solana:2013aba} and~\cite{Casalderrey-Solana:2013sxa} to discuss here seem to be transient effects occurring when the longitudinal profile of the projectiles is sufficiently thin compared to the emergent master scale being the effective temperature at the moment of applicability of hydrodynamics in the centre of the fireball. In this context, the main message from Ref.~\cite{Casalderrey-Solana:2013sxa} is that such $h_{\pm}(x^{\pm})$ can contain substructure but this substructure does not play any significant role for the way centre of the fireball look like. Fig.~\ref{fig.shocks} depicts the relevant collision of the two Gaussian shocks from Ref.~\cite{Casalderrey-Solana:2013aba} and one can see there that the system is so far from equilibrium that the lab-frame energy density is negative in the vicinity of the future light-cone (the total energy is, of course, positive). Similar effects can be reproduced in free quantum field theories. Furthermore, note that the region of the negative energy density is surrounded by regions in which the energy-density is positive. See, in particular, Ref.~\cite{Ford:1999qv} for a discussion of phenomena of that type. A less-obvious phenomenon is the possibility of the absence of the local rest frame, i.e. in the vicinity of the future light-cone there is a region of spacetime where the matter is extremely far from equilibrium and $\langle T^{\mu \nu} \rangle$ does not possess any time-like eigenvector, i.e. there is no reference frame in which the matter is at rest. Note that this condition serves as a definition of the velocity $u^{\alpha}$ in relativistic hydrodynamics in the absence of conserved charges, so in this region one cannot bring $\langle T^{\mu \nu} \rangle$ to the form~\eqref{eq.TmunuMAIN}. For a further discussion of various aspects of the absence of the local rest frame in these collisions, see Ref.~\cite{Arnold:2014jva}.

Summarizing this lecture, it is not clear (yet) what will be the most interesting lesson stemming from holographic collisions in the context of the QGP physics. One thing that these developments demonstrated is our ability to solve Einstein's equations in AdS in a class of time-dependent settings without making additional symmetry assumptions. This technical development allows, for example, to better understand various aspects of relativistic hydrodynamics and its generalizations, e.g. the question of the relevance of two-derivative corrections in $\Pi^{\mu \nu}$ vs. one-derivative corrections encapsulated by Eq.~\eqref{eq.hydro1}, by comparing solutions of hydrodynamic equations with ab initio calculations in holographic gauge theories. Progress on this front might be relevant for understanding uncertainties from these terms on precision measurements of the QGP shear viscosity, which is one of the goals of the HICs programs in the near future, see, e.g., Ref~\cite{Shen:2015msa}. Another development on this front is the question how small a system can be at fixed average energy density to be amenable for hydrodynamic treatment, see in particular Ref.~\cite{Chesler:2016ceu}. This brings us directly to the next lecture, i.e. the physics of hydrodynamization.

\section{Lecture 3: hydrodynamization \label{sec.lec3}}

The most important lessons from holography are those which contain some degree of universality. The chief reason is, of course, the fact that the holographic gauge theories to the best of our knowledge are \emph{not} exactly the ones encountered in nature, in particular ${\cal N} = 4$ SYM and its more elaborate cousins are not QCD. In the context of the near-equilibrium QGP physics, the lesson number one is certainly the universality\footnote{Note at the same time that the value $\frac{1}{4 \pi}$ is \emph{not} the exact lower bound for quantum field theories~\cite{Buchel:2008vz}.} of the shear viscosity, $\frac{\eta}{s} = \frac{1}{4 \pi}$~\cite{Policastro:2001yc}, in all holographic gauge theories in the supergravity approximation~\cite{Buchel:2003tz}. When it comes to the far-from-equilibrium physics discussed in these lectures, Eq.~\eqref{eq.thydroT} being an empirical and qualitative observation in numerical holography does not have such solid foundations as $\frac{\eta}{s} = \frac{1}{4 \pi}$. The question then is whether the studies of relaxation process at strong coupling has led to a truly universal outcome of potentially lasting effect. The aim of this lecture is to demonstrate that understanding the approach to hydrodynamics ab initio using holography in a class of quantum field theories and making the distinction between local thermalization and hydrodynamization with all associated baggage might be just such a result.

The key idea behind this lecture is rather simple: the energy-momentum tensor is going to take the hydrodynamic form such as Eq.~\eqref{eq.hydro1} when the degrees of freedom not encapsulated by the hydrodynamic ansatz become negligible. In the context of holography (and most likely in the general context as well), this way of thinking about the applicability of hydrodynamics was formulated in Ref.~\cite{Chesler:2009cy}. Note that by phrasing the condition for hydrodynamization in this rather intuitive way we do not require gradients to be small, but whether they can be large and how it depends on a setup of interest needed to be checked explicitly with an ab initio calculation. This is the origin of hydrodynamization $\neq$ local thermalization, since in the latter case one requires that $\langle T^{\mu \nu} \rangle$ to be close to its equilibrium form and gradient corrections such as viscosity terms given by Eq.~\eqref{eq.hydro1} take us away from it.

In this lecture we will discuss hydrodynamization $\neq$ local thermalization using a particularly clean example of an expanding plasma system given by the boost-invariant flow with no transversal expansion and without external sources (as opposed to Ref.~\cite{Chesler:2009cy}). One can think of it as of a crude model for what happens in the centre of the future light cone in planar shock waves collision discussed in previous section and depicted on Fig.~\ref{fig.shocks}. Instead of the lab-frame time $x^{0}$ and the lab-frame expansion axis coordinate $x^{1}$ we will use proper time $\tau$ and (spacetime) rapidity coordinate $y$ defined by
\be
x^{0} = \tau \, \cosh{y} \quad \mathrm{and} \quad x^{1} = \tau \, \sinh{y}.
\ee
Right at centre of the collision (and also at the centre of the resulting matter), $x^{1} = 0$ or equivalently $y = 0$, $\tau \equiv x^{0}$. So far we did nothing, just passed to  curvilinear coordinates in which the Minkowski spacetime looks the following
\be
\label{eq.metricbif}
ds^{2} = -d \tau^{2} + \tau^{2} \, dy^{2} + \left( d x^{2} \right)^{2} + \left( d x^{3} \right)^{2},
\ee
but the key idea now will be to impose the condition of invariance under all the boosts along the expansion ($x^{1}$) axis. Since longitudinal boosts preserve $\tau$ and shift $y$, $y \rightarrow y + \Delta y$, in the $\tau$ - $y$ coordinates the condition of the boost-invariance implies that all the components of $\langle T^{\mu \nu} \rangle$ do not depend on $y$. One can seek for phenomenological justifications of this assumption/approximation~\cite{Bjorken:1982qr}, but for us the boost-invariance will only act as a technical simplification in which $\langle T^{\mu \nu} \rangle$ depends only on a single variable $\tau$ yet gives rise to the hydrodynamic tail in a setup resembling an actual HIC.

For the boost-invariant flow with all the symmetries imposed the most general energy-momentum tensor takes the form
\bea
&&\langle T^{\tau \tau} \rangle = {\cal E}(\tau), \quad \langle T^{y y} \rangle = \frac{1}{\tau^{2}} \left\{ \frac{1}{3} {\cal E}(\tau) - \frac{2}{3} \Delta {\cal P} (\tau)\right\} \nonumber \\
&&\quad \quad \mathrm{and} \quad \langle T^{2 2} \rangle = \langle T^{3 3} \rangle = \frac{1}{3} {\cal E}(\tau) + \frac{1}{3} \Delta {\cal P} (\tau)
\eea
with all the other components vanishing. Similarly to our analysis of homogeneous isotropization, also here we want to invariantly parametrize the deviations from equilibrium. By comparing to Eqs.~\eqref{eq.TmunuMAIN} and~\eqref{eq.hydro0}, we immediately see that these deviations are captured by $\Delta {\cal P}(\tau)$. The key difference with the homogeneous isotropization is that now there is a nontrivial hydrodynamic tail, i.e. $\Delta {\cal P}(\tau)$ obtains contributions in the gradient expansion starting already with the viscous term~\eqref{eq.hydro1}. The simplest measure of deviations from local equilibrium is the ratio of the pressure anisotropy to what would be the equilibrium pressure, i.e. $\frac{\Delta {\cal P}(\tau)}{{\cal E}(\tau)/3}$. We want to view this quantity as a function not of the proper time $\tau$, but rather as a function of dimensionless variable measuring proper time in units of what would be the corresponding effective temperature
\be
\label{eq.wdef}
w = \tau \, T(\tau).
\ee
The reason for using this particular dimensionless variable stems from the results discussed in the first lecture in Sec.~\ref{sec.lec1}, which suggest to view $T(\tau)$ as a fundamental non-equilibrium timescale at the hydrodynamic threshold\footnote{Note that $\frac{\Delta {\cal P}(\tau)}{{\cal E}(\tau)/3}$ is not always single valued when expressed as a function of the $w$ variable, but we will not be bothered with it.}. Now, evaluating Eq.~\eqref{eq.hydro1} in the boost-invariant setup of interest characterized by\footnote{The non-zero gradients come from Christoffel symbols of the metric~\eqref{eq.metricbif}. In particular, this is why we used covariant derivatives in Eq.~\eqref{eq.sigma}.}
\be
u^{\mu} \partial_{\mu} = \partial_{\tau}
\ee
one finds that the gradient expansion corresponds to  the large-$w$ expansion of $\frac{\Delta {\cal P}}{{\cal E}/3}$ with the first three terms being
\be
\label{eq.DeltaPgrads}
\frac{\Delta {\cal P}}{{\cal E}/3} = \frac{0.64}{w} + \frac{0.02}{w^{2}} + \frac{0.01}{w^{3}} + \ldots
\ee
The above measure of deviations from equilibrium, at least naively, does not contain any free parameters and hence we expect it to act as an approximate universal solution of this setup at sufficiently late times. Also, note at this point that the two- (the $\frac{1}{w^{2}}$ term) and three-derivative (the $\frac{1}{w^{3}}$ term) corrections are much smaller than the viscous correction (the $\frac{1}{w}$ term), but this might well be the artifact of considering an extremely symmetric flow. In the following, due to historical reasons, instead of $\frac{\Delta {\cal P}}{{\cal E}/3}$ we will very often use
\be
\label{eq.fdef}
f(w) \equiv \frac{2}{3} + \frac{1}{18} \cdot \frac{\Delta {\cal P}}{{\cal E}/3}.
\ee
To add to the footnote~13, one can think of the phase space for the boost-invariant evolution in holography to be spanned by $w$, $f(w)$ and all its derivatives. This follows from the presence of infinitely many QNMs contributing to Eq.~\eqref{eq.TmunuMODES}, see also Eq.~\eqref{eq.transseries} later in this lecture.

\begin{figure}
\begin{center}
\includegraphics[width=9cm]{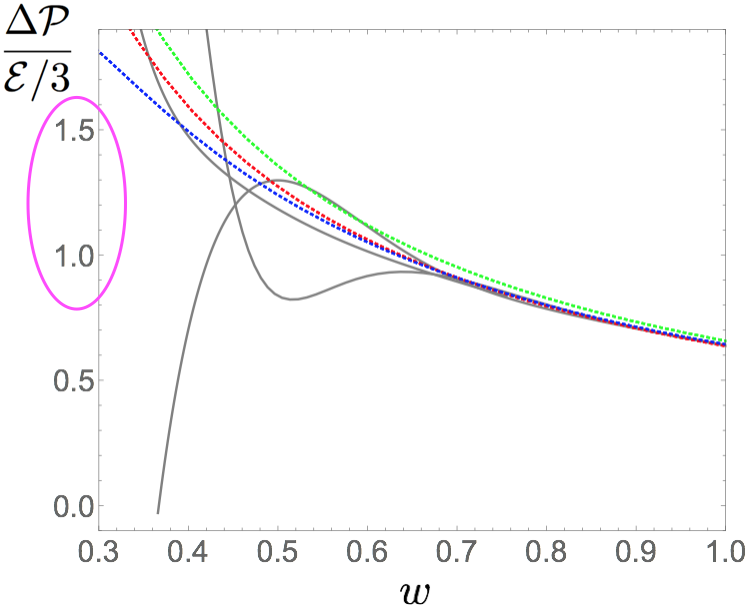}
\end{center}
\caption{The plot of normalized pressure anisotropy as a function of the $w$ time variable defined by Eq.~\eqref{eq.wdef} for three different ab initio solutions of non-equilibrium dynamics of ${\cal N} = 4$ SYM (grey curves) vs. prediction given by truncated gradient expansion at the first (red), second (green) and third (blue) order in derivatives, see also Eq.~\eqref{eq.DeltaPgrads}. Note that the hydrodynamic predictions match extremely well the results of ab initio studies already from $w \approx 0.7$, which is in agreement with Eq.~\eqref{eq.thydroT}. Quite remarkably, this leads to an enormous anisotropy in the hydrodynamic phase driven in this setup by the shear viscosity contribution. The plot adapted from Ref.~\cite{Heller:2011ju} is a particularly clear demonstration that hydrodynamization $\neq$ local thermalization, which was noticed earlier in Refs.~\cite{Chesler:2009cy,Chesler:2010bi}.
\label{fig.hydrodynamization}}
\end{figure}

\begin{figure}
\begin{center}
\includegraphics[width=9cm]{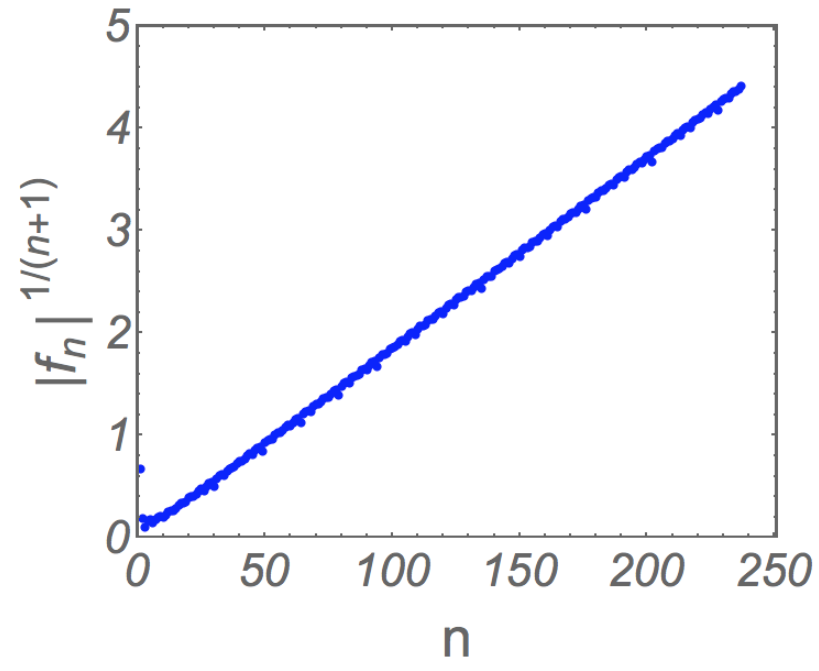}
\end{center}
\caption{The first 241 coefficients in the large-$w$ expansion of combined Eqs.~\eqref{eq.DeltaPgrads} and~\eqref{eq.fdef}. The linear behaviour of $|f_{n}|^{1/(n+1)}$ implies that $f_{n} \sim n!$, which means that the hydrodynamic gradient expansion does not converge for any finite $w$. The plot is adopted from Ref.~\cite{Heller:2013fn}.
\label{fig.divergence}}
\end{figure}

We do not know how many terms in the gradient expansion~\eqref{eq.DeltaPgrads} are optimal (more on this later), so let us take what we have so far, i.e. the three terms in Eq.~\eqref{eq.DeltaPgrads}, and plot them against data from ab initio simulations starting from a bunch of non-equilibrium states. The results of this comparison are depicted on Fig.~\ref{fig.hydrodynamization} which very neatly illustrates the following key notions:
\begin{itemize}
\item At the hydrodynamic threshold the pressure anisotropy can be enormous, between approximately 75\% and 150\% of what would be the equilibrium pressure for considered non-equilibrium states. This is precisely the statement of the hydrodynamization, i.e. the applicability of hydrodynamics does not require local isotropization of the energy-momentum tensor.
\item Including gradient corrections is essential in order to match the ab initio solution when it is in the hydrodynamic regime.
\item Note that whereas viscous hydrodynamics works already at $w = O(1)$, see Eq.~\eqref{eq.thydroT}, the isotropization occurs at values of $w$ of the order of magnitude larger. This directly translates to the order of magnitude larger timescale of isotropization than the timescale needed for the hydrodynamization. 
\end{itemize}
How comes then that the hydrodynamics gradient expansion truncated basically at the lowest nontrivial order does such a remarkable job in matching ab initio solutions in what would be at least naively regarded as a rather extreme regime? This turns out to be rather easy to understand if one recalls that $\frac{\Delta {\cal P}}{{\cal E}/3}$ contains not only the contributions from the gradient expansion encapsulated by Eq.~\eqref{eq.DeltaPgrads}, but also the contributions from transient QNMs of the form
\be
f(w) = \ldots + e^{-\# \, w} \left( \ldots \right) + \ldots
\ee
where the exponential should be compared with Eq.~\eqref{eq.TmunuMODES} and the factor of $w$ comes from the QNMs frequencies being temperature-dependent, see Refs.~\cite{Janik:2006gp,Heller:2013fn,Heller:2014wfa} for an extensive discussion of this point. Without working out the details, we already see that the exponentially suppressed terms are part of the most general solution yet they are not included in the hydrodynamic gradient expansion~\eqref{eq.DeltaPgrads}. This necessarily implies that the hydrodynamic gradient expansion cannot converge for any finite value of $w$, i.e. the hydrodynamic gradient expansion is a divergent series. In Ref.~\cite{Heller:2013fn} we used numerical holography to generate the first several hundred terms in Eq.~\eqref{eq.DeltaPgrads} and explicitly verified this conclusion, see Fig.~\ref{fig.divergence}.

Before we discuss how to properly think about the hydrodynamic gradient expansion in the context of transient QNMs, let us pause and try to view everything we learnt in this lecture from a broader perspective. As it needed to be expected, large deviations from the perfect fluid energy-momentum tensor in the hydrodynamic regime occur also for flows with much less symmetry, in particular in the planar shock waves collisions discussed in the previous lecture and their less symmetric generalizations. One also generically expects that the hydrodynamic gradient expansion is divergent in majority of situations (in particular because we always expect to have transient modes of some sort that are never exactly zero). This statement acts then as a mathematical justification for applying truncated hydrodynamic gradient expansion to model dynamics with very large spatio-temporal variations\footnote{In fact, in HICs one does not use truncated gradient expansion but rather effective equations of motion generating correct gradient expansion to a desirable (first or second) order.}. All-in-all, this is a very good phenomenological news since hydrodynamics is successfully applied to HICs under extreme conditions (in particular, for collisions with large fluctuations in the initial energy density in the transversal plane~\cite{Alver:2010dn} or for some of the collisions involving one or two small projectiles~\cite{Bozek:2011if,Werner:2013tya,Werner:2014xoa}) and the results of ab initio studies such as presented in these lectures show that it can nevertheless do very well even if we only know the first or a few lowest nontrivial terms reliably. The opposite situation in which we would be sensitive to very many orders of the hydrodynamic gradient expansion, reminiscent to considering e.g. a geometric series close to the edge of its radius of convergence, would simply imply the loss of predictability. Finally, note that increasing the number of terms in Eq.~\eqref{eq.DeltaPgrads} at a given fixed value of $w$ is not going to lead to a better match with numerics. However, supposing that one keeps the first three and, say, 100 terms, then there is going to be a value of $w=w_{100}$ such that for $w > w_{100}$ truncated series with 100 terms works better than the series truncated to the first three terms. Of course, in practice we want to be able to cover as wide range of $w$ as possible, hence we use the ansatz with only the first few terms included~\eqref{eq.DeltaPgrads}. The latter is also a desirable feature from the point of view of predictability.

\begin{figure}
\begin{center}
\includegraphics[width=9cm]{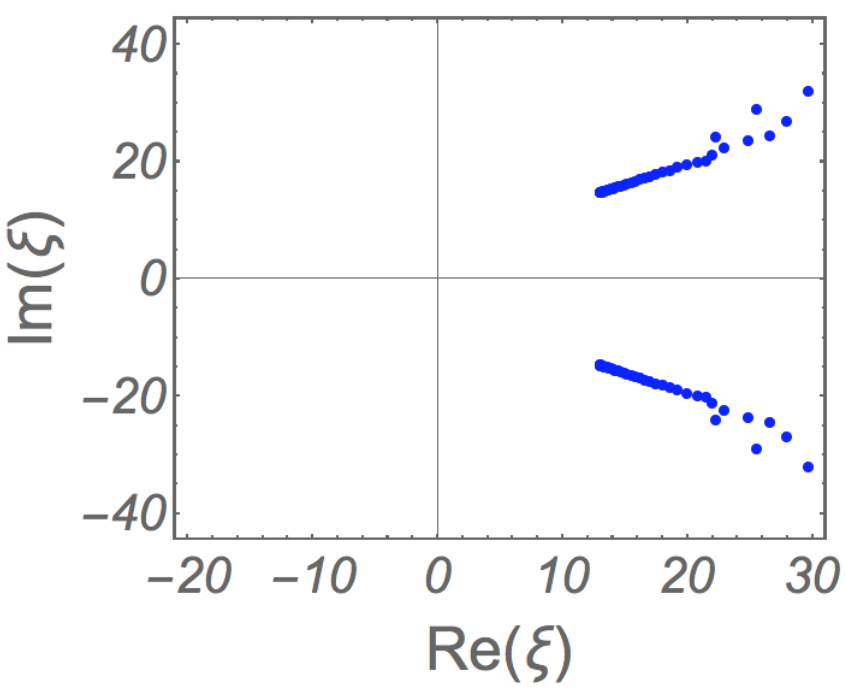}
\end{center}
\caption{Singularities of the approximate analytic continuation of the Borel transform~\eqref{eq.fB} of the large-order hydrodynamic series from Fig.~\ref{fig.divergence}. The approximate analytic continuation is performed using symmetric Pad{\'e} approximants and we plot zeros of the denominator which are not simultaneously zeros of the numerator. The leading singularities are two symmetric branch cuts (Pad{\'e} approximants capture branch cuts as condensations of single poles) which start at $3/2 \, i$ times the frequency of the least damped transient QNM with the accuracy of five decimal places, see Ref.~\cite{Heller:2013fn} for details. Different ways of aligning the integration contour in Eq.~\eqref{eq.Borelsum} with respect to these singularities lead to the idea that the most general ansatz for $f(w)$ has the transseries form, see Eq.~\eqref{eq.transseries}. In particular, we expect many more singularities to appear associated with all transient QNMs and integer multiples of their frequencies, see Ref.~\cite{Buchel:2016cbj}, but we adopted too crude approximation of the analytic continuation to see them. The plot is adapted from Ref.~\cite{Heller:2013fn}.
\label{fig.singpade}}
\end{figure}

Moving on to understanding the interplay between hydrodynamic gradient expansion and transient QNMs, let us recall that the standard method in dealing with divergent series is the Borel transform
\be
\label{eq.fB}
f_B(\xi) = \sum_{n=0}^\infty \frac{f_n}{n!} \, \xi^{n}
\ee
and Borel summation
\be
\label{eq.Borelsum}
f_{I-B}(w) = \frac{1}{w} \int_{\cal C} d\xi \, e^{-\xi/w}\, f_{B}(\xi),
\ee
where the contour ${\cal C}$ connects $0$ with $\infty$. The Borel transform~\eqref{eq.fB} from a series with zero radius of convergence generates a series with a finite radius of convergence dictated by the presence of singularities in its analytic continuation, see Fig.~\ref{fig.singpade} for the approximate structure of the leading singularity for the hydrodynamic series~\eqref{eq.DeltaPgrads}. The latter implies that the Borel summation given by Eq.~\eqref{eq.Borelsum} is an ambiguous procedure and depends on the alignment of the integration contour with respect to encountered singularities. We should not worry that the sum of the divergent series is ill-defined, because we know that hydrodynamic gradient expansion does not appear only by itself -- it is accompanied by the transient modes. In fact, in Ref.~\cite{Heller:2013fn} we showed that the difference between various contours is, at large $w$, given by the leading order contribution from the least damped transient QNM in ${\cal N} = 4$ SYM.

Combining these various pieces of evidence with the studies of a toy-model in Ref.~\cite{Heller:2015dha} and holographic studies of another flow in Ref.~\cite{Buchel:2016cbj} makes us \emph{expect} that the \emph{full} ansatz for $f(w)$, denoted here $f_{ts}$, takes the form
\bea
\label{eq.transseries}
f_{ts}(w) = && \sum_{i = 0}^{\infty}  f_{i} \, w^{-i} + \sum_{j = 1}^{\infty} e^{-\frac{3}{2} \hat{\omega}_{j} w} w^{\alpha_{j}} \sum_{i = 0}^{\infty} g_{j,i} \, w^{-i} + \nonumber \\ &&+ \sum_{j = 1}^{\infty} \sum_{n = 1}^{\infty} e^{-\frac{3}{2} \hat{\omega}_{j} w} e^{-\frac{3}{2} \hat{\omega}_{n} w} w^{\alpha_{j}} w^{\alpha_{n}} \sum_{i = 0}^{\infty} g_{j,n,i} \, w^{-i} + \ldots
\eea
This expression calls for several comments:
\begin{itemize}
\item The first sum is the divergent hydrodynamic gradient expansion.
\item The second sum is the sum (the $j$ index) over contributions from all transient QNMs whose frequencies $\hat{\omega}_{j} \equiv \frac{1}{T} \, \omega_{mode\, j}(k = 0)$ are evaluated at vanishing momentum and over (also divergent) gradient corrections (the $i$ index) to the leading order exponential behaviour. We also need to take into account the power-law exponent $\alpha_{j}$ whose origin is discussed in Refs.~\cite{Heller:2013fn,Heller:2015dha}. One can think of the sum over the $i$ index and the power-law exponent $\alpha_{j}$ as of arising from the interactions of the transient QNMs with hydrodynamic background, i.e. finding QNMs not in the AdS-Schwarzschild background of Eq.~\eqref{eq.bh} but rather in the fluid-gravity duality solution keeping arbitrary many derivatives of $T$ and $u^{\mu}$.
\item The third sum (the indices $j$ and $n$) is a sum over the quadratic terms arising from the leading order interactions between the transient QNMs having to do with the nonlinear nature of Einstein's equations. The sum over the $i$ index is expected to be divergent. The terms in the ellipsis contain interactions between $3$ and more modes and associated expansions in powers of $\frac{1}{w}$, which we again expect to be divergent.
\item Every sum of the $i$ index is understood to be a Borel summation of the analytic continuation of the Borel transform of the original expression with the associated discrete choice of integration contour in the corresponding Eq.~\eqref{eq.Borelsum}.
\item The series contains free parameters being $f_{j,0}$ that we (partly, see the next point) associate with integration constants having to do with the choice of non-equilibrium state we consider. You should think of them to a certain extent as analogues of the term $ \left[ \mathrm{res}_{\omega = \omega_{mode} (k)} G_{R} (\omega, \, \vec{k}) \cdot \delta g(\omega_{mode}(k), \vec{k}) \right]\,^{\mu \nu}$ from Eq.~\eqref{eq.TmunuMODES} evaluated at vanishing momentum, i.e. for $\vec{k} = 0$.
\item Each divergent series in Eq.~\eqref{eq.transseries} needs to be resummed. The choice of the integration contour for the $f_{i}$ series is expected to lead to an ambiguity that can be exactly cancelled by ambiguous \emph{complex} parts of free parameters $f_{j, 0}$. Higher order parts of Eq.~\eqref{eq.transseries} are expected to follow the same pattern, but now the choice of contour made for the $f_{i}$ resummation and the associated choice of $f_{j, 0}$ will determine the choice of contours for resummation of the each of the $f_{j,n,i}$ series at fixed $j$ and $n$.
\item Expression given by Eq.~\eqref{eq.transseries} is an example of a transseries and special relations between parts with different exponentials (including the hydrodynamic gradient expansion) are a manifestation of the so-called resurgence.
\item All in all, Eq.~\eqref{eq.transseries} as a whole is well-defined at least for some range of $w$ around $w = \infty$ and is specified by an infinite set of numbers that correspond in a complicated way to a choice of initial conditions in holography. 
\end{itemize}
Having said all this, it is interesting to draw an analogy with quantum mechanical systems (e.g. anharmonic oscillator with quartic interactions with the coupling constant $g$) in which a given $f(w)$ corresponds to the energy of, say, the ground state, expansion in $1/w$ corresponds to the perturbative expansion and QNMs correspond to non-perturbative effects (instantons in this setup). In fact, applications of resurgence and transseries techniques to quantum mechanical theories is a subject of a significant contemporary interest, see e.g. Refs.~\cite{Aniceto:2011nu,Argyres:2012vv,Argyres:2012ka,Dunne:2012ae,Aniceto:2013fka,Cherman:2014ofa}.

The take-home message from this lecture is that the hydrodynamic gradient expansion must diverge because it does \emph{not} capture transient excitations of microscopic theories. Note that the studies reported here are holographic, but the lesson seems general and, in fact, was recently confirmed also within the kinetic theory~\cite{Denicol:2016bjh,Heller:2016rtz}. The interesting thing with divergent series is that, truncated, they are known to work remarkably well also for what one would naively conclude to be large values of their arguments. This is precisely what we observed in this lecture: the gradient expansion truncated at the first nontrivial term can work very well also when the pressure anisotropy is 150\% times bigger than what would be the equilibrium pressure.  Of course, the hydrodynamization is not directly about anisotropies but rather about the size of gradient contributions to the energy-momentum tensor at the hydrodynamic threshold and indeed it has been confirmed as such in less symmetric setup. In the broader picture, presented studies indicate possible applicability of relatively simple (first few terms in the gradient expansion) hydrodynamic theories in regimes of large gradients which are omnipresent in phenomenological hydrodynamic modelling of HICs. This is certainly a very good news from the point of view of phenomenological applications, but its interesting theoretical spin-off, see in particular recent Ref.~\cite{Romatschke:2016hle}, is that the hydrodynamic evolution in HICs does not necessarily always imply experimental access to equilibrated QGP. It remains to be seen whether this observation will have bearing experimental implications, but it is really pleasing to see that it came from the string theory research.

\section{Summary \label{sec.summary}}

If there is the single most important take home message from these lectures, it is that hydrodynamic gradient expansion does not converge and as a result simple theories of hydrodynamics can work very well also outside their naive regime of validity, i.e. when gradient corrections are large. The role of numerical holography was to provide unambiguous evidence for this statement by performing ab initio calculations of time evolution of excited states in strongly-coupled gauge theories. The question of the impact~of
\begin{center}
hydrodynamization $\neq$ local thermalization
\end{center}
on HICs is not settled yet, but the fact that it is so robust makes it one of not-so-many truly universal lessons stemming from the applications of holography. In the worst case scenario, the phenomenon of hydrodynamization can provide partial justification for the applicability of relatively simple (first few terms in the gradient expansion are matched to the microscopic model) hydrodynamic theories in regimes of large gradients which are omnipresent in phenomenological hydrodynamic modelling of HICs, but more studies are needed.

\section{Outlook}

Let us notice that, as it often happens in physics, the phenomenon of hydrodynamization ($\neq$ local thermalization) could have been deduced soon after the first papers on the role of QNMs in holography such as Ref.~\cite{Kovtun:2005ev}. In order to see it\footnote{The argument centred around Eq.~\eqref{eq.Hong} was recently suggested to the author by Hong Liu.} let us consider a toy-model of a hydrodynamic mode contribution to Eq.~\eqref{eq.TmunuMODES} with the term $\left[ \mathrm{res}_{\omega = \omega_{hydro} (k)} G_{R} (\omega, \, \vec{k}) \cdot \delta g(\omega_{hydro}(k), \vec{k}) \right]\,^{\mu \nu}$ replaced by a Gaussian and the integral over momenta taken for simplicity to be one-dimensional:
\be
\label{eq.Hong}
\langle T^{\mu \nu}\rangle_{toy} \sim \int_{-\infty}^{\infty} dk \, e^{i \, k \, x + i \, \omega(k) \, t} e^{-k^{2}} + \ldots
\ee 
In the equation above the ellipsis denotes the contribution from transient modes that we will not be interested in. Now assume that $\omega(k)$ expanded around $k = 0$ (this is the hydrodynamic gradient expansion in the Fourier space) has a non-zero radius of convergence which we know for sure at least in some cases~\cite{Baier:2007ix}. If we now first expand the frequency $\omega(k)$ in $k$, which for a sound wave would look the following
\be
e^{i \, \omega(k) \, t} = e^{\pm i \, \frac{1}{\sqrt{3}} k \, t} \left(1 + \sum_{n = 1}^{\infty} s_{n} \, k^{n}\right),
\ee 
and integrate it over momenta term-by-term, i.e. integrate each of the $s_{n}$ contributions separately, we will end up integrating $k^{n}$ weighted by an exponential which naturally leads to the appearance of $n!$. This indicates that the inverse Fourier transform will return the divergent series and as a result, even in linearized hydrodynamics the gradient expansion in \emph{real space} seems to be divergent. Now when it comes to the hydrodynamization, Eq.~\eqref{eq.DeltaPgrads} (in a slightly different parametrization) was already known in 2007. Since QNMs obey Eq.~\eqref{eq.thydroT}, inserting it into Eq.~\eqref{eq.DeltaPgrads} and keeping only the first few terms (since now we would expect it not to converge!) we arrive at very large pressure anisotropy already for~$w = 1$. From this perspective, the nontrivial input of the ab initio studies in the nonlinear regime such as discussed in these lectures Refs.~\cite{Heller:2011ju,Heller:2012je,Heller:2012km,Heller:2013oxa,Casalderrey-Solana:2013aba,Casalderrey-Solana:2013sxa} is that Eq.~\eqref{eq.thydroT} holds also when deviations from hydrodynamics are large in amplitude.

It is also important to stress that hydrodynamization was reported here only in the simplest setup possible (boost-invariant flow in a CFT) in which the shear viscosity alone was capturing most of the physics. One may expect that there are other flows for which some of the second order terms would give significant contributions. Following this logic, it is also natural to expect that, e.g., when the conformal symmetry is broken then the term coming from nonzero bulk viscosity, see Eq.~\eqref{eq.hydro1}, can be sizeable. Indeed, recently this phenomenon was observed in holographic studies reported in Ref.~\cite{Attems:2016tby}. In the presence of anomalies~\cite{Son:2009tf} or for magnetohydrodynamics~\cite{Huang:2011dc} there are interesting other terms at the first order in derivatives (for triangle anomalies they contribute to the current) and by the same logic one may expect them also to be quite sizeable at the hydrodynamic threshold but the author is unaware of such results.

On a more formal side, note that we still do not understand the mathematical reason behind the divergent character of the hydrodynamics gradient expansion, i.e. whether it comes from individual gradient terms giving very large contribution to, say, the boost-invariant flow at each order or rather a very large number of terms which add up. Some progress on this issue can be achieved by studying the hydrodynamic gradient expansion without any symmetry assumptions in various theories of relativistic hydrodynamics generalizing the analysis in Refs.~\cite{Heller:2015dha,Basar:2015ava,Aniceto:2015mto}. Also, the transseries ansatz for holographic theories given by Eq.~\eqref{eq.transseries} with most of its resurgent properties conjectured is based on what we learnt from Refs.~\cite{Heller:2015dha,Basar:2015ava,Aniceto:2015mto,Buchel:2016cbj} and it would be really interesting to understand if it is indeed correct. A particularly interesting aspect of it is a sum over all the QNMs which might not form a basis of functions in the standard sense~\cite{Warnick:2013hba}\footnote{The author would like to thank Luis Lehner for bringing this issue up.}. 

It should be re-stressed here that the conclusions coming from the studies of pure Einstein gravity with negative cosmological constant will certainly be somewhat altered by including higher derivative corrections to the gravitational action. In order to trust performed calculations, one should treat them as small corrections to all the quantities discussed here so these changes are not expected to be big. Nevertheless, it would be really interesting to comprehensively understand associated trends, e.g. if a given correction lowers the value of the ratio $\frac{\eta}{s}$, how does it affect hydrodynamization times and how does it influence the pressure anisotropy at the hydrodynamic threshold? Preliminary works in these directions include Refs.~\cite{Grozdanov:2016vgg,Grozdanov:2016zjj,Andrade:2016rln}. In certain cases one can consider, as toy-models, situations in which the associated correction is sizable, see Ref.~\cite{Grozdanov:2016vgg} for such a study. In these cases, it is possible to change in a quantitative way the structure of singularities of retarded Green's functions of the energy-momentum tensor, which as a result leads to different lessons than those presented in the lectures. Although this particular direction is certainly very interesting, by pursuing it one looses the virtue of performing ab initio studies of gauge theories which has been the strongest motivation for using holography in the context of HICs.

Finally, note that applied holography, including numerical holography, delivered a number of other very interesting results at the intersection with HICs that were not discussed in these lectures, in particular having to do with jets. On this front, Ref.~\cite{Chesler:2015lsa} overviews some of those fascinating developments and interested reader is invited to consult it. Note also that even when it comes to the question of the dynamics of holographic QGPs, not much is known about nonlocal correlations. One reason for it is that many groups use for that purpose the so-called AdS-Vaidya spacetime as a gravitational model of equilibration and it lacks most of the crucially interesting features discussed in these lectures. The other is that for calculating correlations one very often uses the approximation in which 2-point correlation functions are given by lengths of spatial geodesics anchored at the boundary, see Ref.~\cite{Balasubramanian:2011ur} for early comprehensive studies of correlations in the geodesic approximation and Ref.~\cite{Ecker:2016thn} for such a study in the colliding shock-waves background discussed in Sec.~\ref{sec.lec2}. To the best of the author's knowledge the only articles to date not adopting any of those simplifications are Refs.~\cite{CaronHuot:2011dr,Chesler:2011ds,Chesler:2012zk}. Certainly further progress is needed on this front and some interesting steps in these directions were taken by Philip Kleinert and his collaborator and were presented by Philip in Zakopane in a talk based on Refs.~\cite{Keranen:2014lna,Keranen:2015mqc}. See also Ref.~\cite{David:2015xqa} for related developments.

Another fascinating topic not covered here is the physics of equilibration processes (or lack of thereof) for strongly-coupled gauge theories on spheres, e.g. ${\cal N} = 4$ on $R \times S^{3}$, better known under the name of the ``AdS (in)stability problem''~\cite{Bizon:2011gg,Maliborski:2013jca,Balasubramanian:2014cja,Craps:2015jma}. Also along the lines of phenomena with multiple scales, studies of turbulence in holographic quantum field theories~\cite{Adams:2012pj,Carrasco:2012nf,Adams:2013vsa} and possible applications to astrophysical black holes~\cite{Green:2013zba,Yang:2014tla} certainly deserve further attention. On this front, it would be very interesting to understand if in (1+3)-dimensional relativistic hydrodynamics the cascade to the UV dissipates through viscosity or if the transient QNMs are needed.

Last but not least, let us emphasize one again that numerical studies discussed in these lectures were all characterized by uniform event horizon and some of the most challenging open problems in numerical holography have to do with situation in which either the horizon is formed as a result of an (possibly very long) evolution process (this occurs in the ``AdS (in)stability problem'') or when two horizons merge into a single black hole (imagine collisions of black holes which unlike gravitational shock-waves given by Eq.~\eqref{eq.shocks} would correspond to projectiles with adjustable $\gamma$-factors).

\acknowledgments

The author would like to thank his collaborators, the organizers of the Summer School on String Theory and Holography in Lisbon (Portugal, 2014) and the 56$^\mathrm{th}$ Cracow School for Theoretical Physics ``A Panorama of Holography'' in Zakopane (Poland, 2016) for the opportunity to present these lectures and participants of these meetings for interesting questions and discussions. Research at Perimeter Institute is supported by the Government of Canada through the Department of Innovation, Science and Economic Development and by the Province of Ontario through the Ministry
of Research \& Innovation.

\bibliography{heller-zakopane-2016}{}

\providecommand{\href}[2]{#2}\begingroup\raggedright\begin{thebibliography}{100}

\bibitem{Maldacena:1997re}
J.~M. Maldacena, ``{The Large N limit of superconformal field theories and
  supergravity},'' {\em Adv.Theor.Math.Phys.} {\bf 2} (1998)  231--252,
\href{http://arxiv.org/abs/hep-th/9711200}{{\tt arXiv:hep-th/9711200
  [hep-th]}}.
%\%CITATION = HEP-TH/9711200;\%\%.

\bibitem{Gubser:1998bc}
S.~Gubser, I.~R. Klebanov, and A.~M. Polyakov, ``{Gauge theory correlators from
  noncritical string theory},''
  \href{http://dx.doi.org/10.1016/S0370-2693(98)00377-3}{{\em Phys.Lett.} {\bf
  B428} (1998)  105--114},
\href{http://arxiv.org/abs/hep-th/9802109}{{\tt arXiv:hep-th/9802109
  [hep-th]}}.
%\%CITATION = HEP-TH/9802109;\%\%.

\bibitem{Witten:1998qj}
E.~Witten, ``{Anti-de Sitter space and holography},'' {\em
  Adv.Theor.Math.Phys.} {\bf 2} (1998)  253--291,
\href{http://arxiv.org/abs/hep-th/9802150}{{\tt arXiv:hep-th/9802150
  [hep-th]}}.
%\%CITATION = HEP-TH/9802150;\%\%.

\bibitem{Gyulassy:2004zy}
M.~Gyulassy and L.~McLerran, ``{New forms of QCD matter discovered at RHIC},''
  \href{http://dx.doi.org/10.1016/j.nuclphysa.2004.10.034}{{\em Nucl. Phys.}
  {\bf A750} (2005)  30--63},
\href{http://arxiv.org/abs/nucl-th/0405013}{{\tt arXiv:nucl-th/0405013
  [nucl-th]}}.
%%CITATION = NUCL-TH/0405013;%%.

\bibitem{Muller:2012zq}
B.~Muller, J.~Schukraft, and B.~Wyslouch, ``{First Results from Pb+Pb
  collisions at the LHC},''
  \href{http://dx.doi.org/10.1146/annurev-nucl-102711-094910}{{\em Ann. Rev.
  Nucl. Part. Sci.} {\bf 62} (2012)  361--386},
\href{http://arxiv.org/abs/1202.3233}{{\tt arXiv:1202.3233 [hep-ex]}}.
%%CITATION = ARXIV:1202.3233;%%.

\bibitem{Shuryak:1988ck}
E.~V. Shuryak, ``{The QCD vacuum, hadrons and the superdense matter},''
{\em World Sci. Lect. Notes Phys.} {\bf 71} (2004)  1--618.
%%CITATION = 00327,71,1;%%.

\bibitem{Wiedemann:2012py}
U.~A. Wiedemann, ``{Introductory Overview of Quark Matter 2012},''
  \href{http://dx.doi.org/10.1016/j.nuclphysa.2013.01.038}{{\em Nucl. Phys.}
  {\bf A904-905} (2013)  3c--10c},
\href{http://arxiv.org/abs/1212.3306}{{\tt arXiv:1212.3306 [hep-ph]}}.
%%CITATION = ARXIV:1212.3306;%%.

\bibitem{Baier:2007ix}
R.~Baier, P.~Romatschke, D.~T. Son, A.~O. Starinets, and M.~A. Stephanov,
  ``{Relativistic viscous hydrodynamics, conformal invariance, and
  holography},'' \href{http://dx.doi.org/10.1088/1126-6708/2008/04/100}{{\em
  JHEP} {\bf 04} (2008)  100},
\href{http://arxiv.org/abs/0712.2451}{{\tt arXiv:0712.2451 [hep-th]}}.
%%CITATION = ARXIV:0712.2451;%%.

\bibitem{vanderSchee:2012qj}
W.~van~der Schee, ``{Holographic thermalization with radial flow},''
  \href{http://dx.doi.org/10.1103/PhysRevD.87.061901}{{\em Phys. Rev.} {\bf
  D87} (2013) no.~6, 061901},
\href{http://arxiv.org/abs/1211.2218}{{\tt arXiv:1211.2218 [hep-th]}}.
%%CITATION = ARXIV:1211.2218;%%.

\bibitem{vanderSchee:2013pia}
W.~van~der Schee, P.~Romatschke, and S.~Pratt, ``{Fully Dynamical Simulation of
  Central Nuclear Collisions},''
  \href{http://dx.doi.org/10.1103/PhysRevLett.111.222302}{{\em Phys. Rev.
  Lett.} {\bf 111} (2013) no.~22, 222302},
\href{http://arxiv.org/abs/1307.2539}{{\tt arXiv:1307.2539}}.
%%CITATION = ARXIV:1307.2539;%%.

\bibitem{Chesler:2015wra}
P.~M. Chesler and L.~G. Yaffe, ``{Holography and off-center collisions of
  localized shock waves},''
  \href{http://dx.doi.org/10.1007/JHEP10(2015)070}{{\em JHEP} {\bf 10} (2015)
  070},
\href{http://arxiv.org/abs/1501.04644}{{\tt arXiv:1501.04644 [hep-th]}}.
%%CITATION = ARXIV:1501.04644;%%.

\bibitem{Chesler:2016ceu}
P.~M. Chesler, ``{How big are the smallest drops of quark-gluon plasma?},''
  \href{http://dx.doi.org/10.1007/JHEP03(2016)146}{{\em JHEP} {\bf 03} (2016)
  146},
\href{http://arxiv.org/abs/1601.01583}{{\tt arXiv:1601.01583 [hep-th]}}.
%%CITATION = ARXIV:1601.01583;%%.

\bibitem{Policastro:2001yc}
G.~Policastro, D.~T. Son, and A.~O. Starinets, ``{The Shear viscosity of
  strongly coupled N=4 supersymmetric Yang-Mills plasma},''
  \href{http://dx.doi.org/10.1103/PhysRevLett.87.081601}{{\em Phys. Rev. Lett.}
  {\bf 87} (2001)  081601},
\href{http://arxiv.org/abs/hep-th/0104066}{{\tt arXiv:hep-th/0104066
  [hep-th]}}.
%%CITATION = HEP-TH/0104066;%%.

\bibitem{Huot:2006ys}
S.~C. Huot, S.~Jeon, and G.~D. Moore, ``{Shear viscosity in weakly coupled N =
  4 super Yang-Mills theory compared to QCD},''
  \href{http://dx.doi.org/10.1103/PhysRevLett.98.172303}{{\em Phys. Rev. Lett.}
  {\bf 98} (2007)  172303},
\href{http://arxiv.org/abs/hep-ph/0608062}{{\tt arXiv:hep-ph/0608062
  [hep-ph]}}.
%%CITATION = HEP-PH/0608062;%%.

\bibitem{Arnold:2007pg}
P.~B. Arnold, ``{Quark-Gluon Plasmas and Thermalization},''
  \href{http://dx.doi.org/10.1142/S021830130700832X}{{\em Int. J. Mod. Phys.}
  {\bf E16} (2007)  2555--2594},
\href{http://arxiv.org/abs/0708.0812}{{\tt arXiv:0708.0812 [hep-ph]}}.
%%CITATION = ARXIV:0708.0812;%%.

\bibitem{Heller:2011ju}
M.~P. Heller, R.~A. Janik, and P.~Witaszczyk, ``{The characteristics of
  thermalization of boost-invariant plasma from holography},''
  \href{http://dx.doi.org/10.1103/PhysRevLett.108.201602}{{\em Phys.Rev.Lett.}
  {\bf 108} (2012)  201602},
\href{http://arxiv.org/abs/1103.3452}{{\tt arXiv:1103.3452 [hep-th]}}.
%%CITATION = ARXIV:1103.3452;%%.

\bibitem{Heller:2012je}
M.~P. Heller, R.~A. Janik, and P.~Witaszczyk, ``{A numerical relativity
  approach to the initial value problem in asymptotically Anti-de Sitter
  spacetime for plasma thermalization - an ADM formulation},''
  \href{http://dx.doi.org/10.1103/PhysRevD.85.126002}{{\em Phys. Rev.} {\bf
  D85} (2012)  126002},
\href{http://arxiv.org/abs/1203.0755}{{\tt arXiv:1203.0755 [hep-th]}}.
%%CITATION = ARXIV:1203.0755;%%.

\bibitem{Heller:2012km}
M.~P. Heller, D.~Mateos, W.~van~der Schee, and D.~Trancanelli, ``{Strong
  Coupling Isotropization of Non-Abelian Plasmas Simplified},''
  \href{http://dx.doi.org/10.1103/PhysRevLett.108.191601}{{\em Phys.Rev.Lett.}
  {\bf 108} (2012)  191601},
\href{http://arxiv.org/abs/1202.0981}{{\tt arXiv:1202.0981 [hep-th]}}.
%%CITATION = ARXIV:1202.0981;%%.

\bibitem{Heller:2013oxa}
M.~P. Heller, D.~Mateos, W.~van~der Schee, and M.~Triana, ``{Holographic
  isotropization linearized},''
  \href{http://dx.doi.org/10.1007/JHEP09(2013)026}{{\em JHEP} {\bf 1309} (2013)
   026},
\href{http://arxiv.org/abs/1304.5172}{{\tt arXiv:1304.5172 [hep-th]}}.
%%CITATION = ARXIV:1304.5172;%%.

\bibitem{Casalderrey-Solana:2013aba}
J.~Casalderrey-Solana, M.~P. Heller, D.~Mateos, and W.~van~der Schee, ``{From
  full stopping to transparency in a holographic model of heavy ion
  collisions},'' \href{http://dx.doi.org/10.1103/PhysRevLett.111.181601}{{\em
  Phys.Rev.Lett.} {\bf 111} (2013)  181601},
\href{http://arxiv.org/abs/1305.4919}{{\tt arXiv:1305.4919 [hep-th]}}.
%%CITATION = ARXIV:1305.4919;%%.

\bibitem{Casalderrey-Solana:2013sxa}
J.~Casalderrey-Solana, M.~P. Heller, D.~Mateos, and W.~van~der Schee,
  ``{Longitudinal Coherence in a Holographic Model of Asymmetric Collisions},''
  \href{http://dx.doi.org/10.1103/PhysRevLett.112.221602}{{\em Phys. Rev.
  Lett.} {\bf 112} (2014) no.~22, 221602},
\href{http://arxiv.org/abs/1312.2956}{{\tt arXiv:1312.2956 [hep-th]}}.
%%CITATION = ARXIV:1312.2956;%%.

\bibitem{Buchel:2015saa}
A.~Buchel, M.~P. Heller, and R.~C. Myers, ``{Equilibration rates in a strongly
  coupled nonconformal quark-gluon plasma},''
  \href{http://dx.doi.org/10.1103/PhysRevLett.114.251601}{{\em Phys. Rev.
  Lett.} {\bf 114} (2015) no.~25, 251601},
\href{http://arxiv.org/abs/1503.07114}{{\tt arXiv:1503.07114 [hep-th]}}.
%%CITATION = ARXIV:1503.07114;%%.

\bibitem{Heller:2013fn}
M.~P. Heller, R.~A. Janik, and P.~Witaszczyk, ``{Hydrodynamic Gradient
  Expansion in Gauge Theory Plasmas},''
  \href{http://dx.doi.org/10.1103/PhysRevLett.110.211602}{{\em Phys.Rev.Lett.}
  {\bf 110} (2013) no.~21, 211602},
\href{http://arxiv.org/abs/1302.0697}{{\tt arXiv:1302.0697 [hep-th]}}.
%%CITATION = ARXIV:1302.0697;%%.

\bibitem{Heller:2015dha}
M.~P. Heller and M.~Spalinski, ``{Hydrodynamics Beyond the Gradient Expansion:
  Resurgence and Resummation},''
  \href{http://dx.doi.org/10.1103/PhysRevLett.115.072501}{{\em Phys. Rev.
  Lett.} {\bf 115} (2015) no.~7, 072501},
  \href{http://arxiv.org/abs/1503.07514}{{\tt arXiv:1503.07514 [hep-th]}}.

\bibitem{Buchel:2016cbj}
A.~Buchel, M.~P. Heller, and J.~Noronha, ``{Beyond adiabatic approximation in
  Big Bang Cosmology: hydrodynamics, resurgence and entropy production in the
  Universe},''
\href{http://arxiv.org/abs/1603.05344}{{\tt arXiv:1603.05344 [hep-th]}}.
%%CITATION = ARXIV:1603.05344;%%.

\bibitem{Kurkela:2015qoa}
A.~Kurkela and Y.~Zhu, ``{Isotropization and hydrodynamization in weakly
  coupled heavy-ion collisions},''
  \href{http://dx.doi.org/10.1103/PhysRevLett.115.182301}{{\em Phys. Rev.
  Lett.} {\bf 115} (2015) no.~18, 182301},
\href{http://arxiv.org/abs/1506.06647}{{\tt arXiv:1506.06647 [hep-ph]}}.
%%CITATION = ARXIV:1506.06647;%%.

\bibitem{Keegan:2015avk}
L.~Keegan, A.~Kurkela, P.~Romatschke, W.~van~der Schee, and Y.~Zhu, ``{Weak and
  strong coupling equilibration in nonabelian gauge theories},''
\href{http://arxiv.org/abs/1512.05347}{{\tt arXiv:1512.05347 [hep-th]}}.
%%CITATION = ARXIV:1512.05347;%%.

\bibitem{Heller:2016rtz}
M.~P. Heller, A.~Kurkela, and M.~Spalinski, ``{Hydrodynamization and transient
  modes of expanding plasma in kinetic theory},''
\href{http://arxiv.org/abs/1609.04803}{{\tt arXiv:1609.04803 [nucl-th]}}.
%%CITATION = ARXIV:1609.04803;%%.

\bibitem{Romatschke:2016hle}
P.~Romatschke, ``{Do nuclear collisions create a locally equilibrated
  quark-gluon plasma?},''
\href{http://arxiv.org/abs/1609.02820}{{\tt arXiv:1609.02820 [nucl-th]}}.
%%CITATION = ARXIV:1609.02820;%%.

\bibitem{Arnold:2002zm}
P.~B. Arnold, G.~D. Moore, and L.~G. Yaffe, ``{Effective kinetic theory for
  high temperature gauge theories},''
  \href{http://dx.doi.org/10.1088/1126-6708/2003/01/030}{{\em JHEP} {\bf 01}
  (2003)  030},
\href{http://arxiv.org/abs/hep-ph/0209353}{{\tt arXiv:hep-ph/0209353
  [hep-ph]}}.
%%CITATION = HEP-PH/0209353;%%.

\bibitem{Gelis:2015gza}
F.~Gelis, ``{Initial state and thermalization in the Color Glass Condensate
  framework},'' \href{http://dx.doi.org/10.1142/S0218301315300088}{{\em Int. J.
  Mod. Phys.} {\bf E24} (2015) no.~10, 1530008},
\href{http://arxiv.org/abs/1508.07974}{{\tt arXiv:1508.07974 [hep-ph]}}.
%%CITATION = ARXIV:1508.07974;%%.

\bibitem{Bhaseen:2012gg}
M.~J. Bhaseen, J.~P. Gauntlett, B.~D. Simons, J.~Sonner, and T.~Wiseman,
  ``{Holographic Superfluids and the Dynamics of Symmetry Breaking},''
  \href{http://dx.doi.org/10.1103/PhysRevLett.110.015301}{{\em Phys. Rev.
  Lett.} {\bf 110} (2013) no.~1, 015301},
\href{http://arxiv.org/abs/1207.4194}{{\tt arXiv:1207.4194 [hep-th]}}.
%%CITATION = ARXIV:1207.4194;%%.

\bibitem{Adams:2012pj}
A.~Adams, P.~M. Chesler, and H.~Liu, ``{Holographic Vortex Liquids and
  Superfluid Turbulence},''
  \href{http://dx.doi.org/10.1126/science.1233529}{{\em Science} {\bf 341}
  (2013)  368--372},
\href{http://arxiv.org/abs/1212.0281}{{\tt arXiv:1212.0281 [hep-th]}}.
%%CITATION = ARXIV:1212.0281;%%.

\bibitem{Sonner:2014tca}
J.~Sonner, A.~del Campo, and W.~H. Zurek, ``{Universal far-from-equilibrium
  Dynamics of a Holographic Superconductor},''
  \href{http://arxiv.org/abs/1406.2329}{{\tt arXiv:1406.2329 [hep-th]}}.
[Nature Commun.6,7406(2015)].
%%CITATION = ARXIV:1406.2329;%%.

\bibitem{Bhattacharyya:2008mz}
S.~Bhattacharyya, R.~Loganayagam, I.~Mandal, S.~Minwalla, and A.~Sharma,
  ``{Conformal Nonlinear Fluid Dynamics from Gravity in Arbitrary
  Dimensions},'' \href{http://dx.doi.org/10.1088/1126-6708/2008/12/116}{{\em
  JHEP} {\bf 12} (2008)  116},
\href{http://arxiv.org/abs/0809.4272}{{\tt arXiv:0809.4272 [hep-th]}}.
%%CITATION = ARXIV:0809.4272;%%.

\bibitem{deHaro:2000vlm}
S.~de~Haro, S.~N. Solodukhin, and K.~Skenderis, ``{Holographic reconstruction
  of space-time and renormalization in the AdS / CFT correspondence},''
  \href{http://dx.doi.org/10.1007/s002200100381}{{\em Commun. Math. Phys.} {\bf
  217} (2001)  595--622},
\href{http://arxiv.org/abs/hep-th/0002230}{{\tt arXiv:hep-th/0002230
  [hep-th]}}.
%%CITATION = HEP-TH/0002230;%%.

\bibitem{Janik:2005zt}
R.~A. Janik and R.~B. Peschanski, ``{Asymptotic perfect fluid dynamics as a
  consequence of Ads/CFT},''
  \href{http://dx.doi.org/10.1103/PhysRevD.73.045013}{{\em Phys. Rev.} {\bf
  D73} (2006)  045013},
\href{http://arxiv.org/abs/hep-th/0512162}{{\tt arXiv:hep-th/0512162
  [hep-th]}}.
%%CITATION = HEP-TH/0512162;%%.

\bibitem{Beuf:2009cx}
G.~Beuf, M.~P. Heller, R.~A. Janik, and R.~Peschanski, ``{Boost-invariant early
  time dynamics from AdS/CFT},''
  \href{http://dx.doi.org/10.1088/1126-6708/2009/10/043}{{\em JHEP} {\bf 10}
  (2009)  043},
\href{http://arxiv.org/abs/0906.4423}{{\tt arXiv:0906.4423 [hep-th]}}.
%%CITATION = ARXIV:0906.4423;%%.

\bibitem{Witten:1998zw}
E.~Witten, ``{Anti-de Sitter space, thermal phase transition, and confinement
  in gauge theories},'' {\em Adv. Theor. Math. Phys.} {\bf 2} (1998)  505--532,
\href{http://arxiv.org/abs/hep-th/9803131}{{\tt arXiv:hep-th/9803131
  [hep-th]}}.
%%CITATION = HEP-TH/9803131;%%.

\bibitem{Bhattacharyya:2008jc}
S.~Bhattacharyya, V.~E. Hubeny, S.~Minwalla, and M.~Rangamani, ``{Nonlinear
  Fluid Dynamics from Gravity},''
  \href{http://dx.doi.org/10.1088/1126-6708/2008/02/045}{{\em JHEP} {\bf 0802}
  (2008)  045},
\href{http://arxiv.org/abs/0712.2456}{{\tt arXiv:0712.2456 [hep-th]}}.
%%CITATION = ARXIV:0712.2456;%%.

\bibitem{Chesler:2008hg}
P.~M. Chesler and L.~G. Yaffe, ``{Horizon formation and far-from-equilibrium
  isotropization in supersymmetric Yang-Mills plasma},''
  \href{http://dx.doi.org/10.1103/PhysRevLett.102.211601}{{\em Phys. Rev.
  Lett.} {\bf 102} (2009)  211601},
\href{http://arxiv.org/abs/0812.2053}{{\tt arXiv:0812.2053 [hep-th]}}.
%%CITATION = ARXIV:0812.2053;%%.

\bibitem{Chesler:2009cy}
P.~M. Chesler and L.~G. Yaffe, ``{Boost invariant flow, black hole formation,
  and far-from-equilibrium dynamics in N = 4 supersymmetric Yang-Mills
  theory},'' \href{http://dx.doi.org/10.1103/PhysRevD.82.026006}{{\em
  Phys.Rev.} {\bf D82} (2010)  026006},
\href{http://arxiv.org/abs/0906.4426}{{\tt arXiv:0906.4426 [hep-th]}}.
%%CITATION = ARXIV:0906.4426;%%.

\bibitem{Chesler:2010bi}
P.~M. Chesler and L.~G. Yaffe, ``{Holography and colliding gravitational shock
  waves in asymptotically AdS$_5$ spacetime},''
  \href{http://dx.doi.org/10.1103/PhysRevLett.106.021601}{{\em Phys.Rev.Lett.}
  {\bf 106} (2011)  021601},
\href{http://arxiv.org/abs/1011.3562}{{\tt arXiv:1011.3562 [hep-th]}}.
%%CITATION = ARXIV:1011.3562;%%.

\bibitem{vanderSchee:2015rta}
W.~van~der Schee and B.~Schenke, ``{Rapidity dependence in holographic heavy
  ion collisions},'' \href{http://dx.doi.org/10.1103/PhysRevC.92.064907}{{\em
  Phys. Rev.} {\bf C92} (2015) no.~6, 064907},
\href{http://arxiv.org/abs/1507.08195}{{\tt arXiv:1507.08195 [nucl-th]}}.
%%CITATION = ARXIV:1507.08195;%%.

\bibitem{Chesler:2013lia}
P.~M. Chesler and L.~G. Yaffe, ``{Numerical solution of gravitational dynamics
  in asymptotically anti-de Sitter spacetimes},''
  \href{http://dx.doi.org/10.1007/JHEP07(2014)086}{{\em JHEP} {\bf 07} (2014)
  086},
\href{http://arxiv.org/abs/1309.1439}{{\tt arXiv:1309.1439 [hep-th]}}.
%%CITATION = ARXIV:1309.1439;%%.

\bibitem{Grandclement:2007sb}
P.~Grandclement and J.~Novak, ``{Spectral methods for numerical relativity},''
  \href{http://dx.doi.org/10.12942/lrr-2009-1}{{\em Living Rev. Rel.} {\bf 12}
  (2009)  1},
\href{http://arxiv.org/abs/0706.2286}{{\tt arXiv:0706.2286 [gr-qc]}}.
%%CITATION = ARXIV:0706.2286;%%.

\bibitem{vanderSchee:2014qwa}
W.~van~der Schee, {\em {Gravitational collisions and the quark-gluon plasma}}.
\newblock PhD thesis, Utrecht U., 2014.
\newblock \href{http://arxiv.org/abs/1407.1849}{{\tt arXiv:1407.1849
  [hep-th]}}.
\newblock
\url{http://inspirehep.net/record/1305268/files/arXiv:1407.1849.pdf}.
\newblock
%%CITATION = ARXIV:1407.1849;%%.

\bibitem{Bantilan:2014sra}
H.~Bantilan and P.~Romatschke, ``{Simulation of Black Hole Collisions in
  Asymptotically Anti–de Sitter Spacetimes},''
  \href{http://dx.doi.org/10.1103/PhysRevLett.114.081601}{{\em Phys. Rev.
  Lett.} {\bf 114} (2015) no.~8, 081601},
\href{http://arxiv.org/abs/1410.4799}{{\tt arXiv:1410.4799 [hep-th]}}.
%%CITATION = ARXIV:1410.4799;%%.

\bibitem{Bjorken:1982qr}
J.~Bjorken, ``{Highly Relativistic Nucleus-Nucleus Collisions: The Central
  Rapidity Region},''
\href{http://dx.doi.org/10.1103/PhysRevD.27.140}{{\em Phys.Rev.} {\bf D27}
  (1983)  140--151}.
%%CITATION = PHRVA,D27,140;%%.

\bibitem{Jankowski:2014lna}
J.~Jankowski, G.~Plewa, and M.~Spalinski, ``{Statistics of thermalization in
  Bjorken Flow},'' \href{http://dx.doi.org/10.1007/JHEP12(2014)105}{{\em JHEP}
  {\bf 12} (2014)  105},
\href{http://arxiv.org/abs/1411.1969}{{\tt arXiv:1411.1969 [hep-th]}}.
%%CITATION = ARXIV:1411.1969;%%.

\bibitem{CasalderreySolana:2011us}
J.~Casalderrey-Solana, H.~Liu, D.~Mateos, K.~Rajagopal, and U.~A. Wiedemann,
  ``{Gauge/String Duality, Hot QCD and Heavy Ion Collisions},''
\href{http://arxiv.org/abs/1101.0618}{{\tt arXiv:1101.0618 [hep-th]}}.
%%CITATION = ARXIV:1101.0618;%%.

\bibitem{DeWolfe:2013cua}
O.~DeWolfe, S.~S. Gubser, C.~Rosen, and D.~Teaney, ``{Heavy ions and string
  theory},'' \href{http://dx.doi.org/10.1016/j.ppnp.2013.11.001}{{\em Prog.
  Part. Nucl. Phys.} {\bf 75} (2014)  86--132},
\href{http://arxiv.org/abs/1304.7794}{{\tt arXiv:1304.7794 [hep-th]}}.
%%CITATION = ARXIV:1304.7794;%%.

\bibitem{Chesler:2015lsa}
P.~M. Chesler and W.~van~der Schee, ``{Early thermalization, hydrodynamics and
  energy loss in AdS/CFT},''
  \href{http://dx.doi.org/10.1142/S0218301315300118}{{\em Int. J. Mod. Phys.}
  {\bf E24} (2015) no.~10, 1530011},
\href{http://arxiv.org/abs/1501.04952}{{\tt arXiv:1501.04952 [nucl-th]}}.
%%CITATION = ARXIV:1501.04952;%%.

\bibitem{Amado:2007yr}
I.~Amado, C.~Hoyos-Badajoz, K.~Landsteiner, and S.~Montero, ``{Residues of
  correlators in the strongly coupled N=4 plasma},''
  \href{http://dx.doi.org/10.1103/PhysRevD.77.065004}{{\em Phys. Rev.} {\bf
  D77} (2008)  065004},
\href{http://arxiv.org/abs/0710.4458}{{\tt arXiv:0710.4458 [hep-th]}}.
%%CITATION = ARXIV:0710.4458;%%.

\bibitem{Kovtun:2005ev}
P.~K. Kovtun and A.~O. Starinets, ``{Quasinormal modes and holography},''
  \href{http://dx.doi.org/10.1103/PhysRevD.72.086009}{{\em Phys.Rev.} {\bf D72}
  (2005)  086009},
\href{http://arxiv.org/abs/hep-th/0506184}{{\tt arXiv:hep-th/0506184
  [hep-th]}}.
%%CITATION = HEP-TH/0506184;%%.

\bibitem{Berti:2009kk}
E.~Berti, V.~Cardoso, and A.~O. Starinets, ``{Quasinormal modes of black holes
  and black branes},''
  \href{http://dx.doi.org/10.1088/0264-9381/26/16/163001}{{\em Class. Quant.
  Grav.} {\bf 26} (2009)  163001},
\href{http://arxiv.org/abs/0905.2975}{{\tt arXiv:0905.2975 [gr-qc]}}.
%%CITATION = ARXIV:0905.2975;%%.

\bibitem{Pilch:2000ue}
K.~Pilch and N.~P. Warner, ``{N=2 supersymmetric RG flows and the IIB
  dilaton},'' \href{http://dx.doi.org/10.1016/S0550-3213(00)00656-8}{{\em Nucl.
  Phys.} {\bf B594} (2001)  209--228},
\href{http://arxiv.org/abs/hep-th/0004063}{{\tt arXiv:hep-th/0004063
  [hep-th]}}.
%%CITATION = HEP-TH/0004063;%%.

\bibitem{Buchel:2003ah}
A.~Buchel and J.~T. Liu, ``{Thermodynamics of the N=2* flow},''
  \href{http://dx.doi.org/10.1088/1126-6708/2003/11/031}{{\em JHEP} {\bf 11}
  (2003)  031},
\href{http://arxiv.org/abs/hep-th/0305064}{{\tt arXiv:hep-th/0305064
  [hep-th]}}.
%%CITATION = HEP-TH/0305064;%%.

\bibitem{Buchel:2007vy}
A.~Buchel, S.~Deakin, P.~Kerner, and J.~T. Liu, ``{Thermodynamics of the N=2*
  strongly coupled plasma},''
  \href{http://dx.doi.org/10.1016/j.nuclphysb.2007.06.019}{{\em Nucl. Phys.}
  {\bf B784} (2007)  72--102},
\href{http://arxiv.org/abs/hep-th/0701142}{{\tt arXiv:hep-th/0701142
  [hep-th]}}.
%%CITATION = HEP-TH/0701142;%%.

\bibitem{Janik:2015waa}
R.~A. Janik, G.~Plewa, H.~Soltanpanahi, and M.~Spalinski, ``{Linearized
  nonequilibrium dynamics in nonconformal plasma},''
  \href{http://dx.doi.org/10.1103/PhysRevD.91.126013}{{\em Phys. Rev.} {\bf
  D91} (2015) no.~12, 126013},
\href{http://arxiv.org/abs/1503.07149}{{\tt arXiv:1503.07149 [hep-th]}}.
%%CITATION = ARXIV:1503.07149;%%.

\bibitem{Fuini:2015hba}
J.~F. Fuini and L.~G. Yaffe, ``{Far-from-equilibrium dynamics of a strongly
  coupled non-Abelian plasma with non-zero charge density or external magnetic
  field},'' \href{http://dx.doi.org/10.1007/JHEP07(2015)116}{{\em JHEP} {\bf
  07} (2015)  116},
\href{http://arxiv.org/abs/1503.07148}{{\tt arXiv:1503.07148 [hep-th]}}.
%%CITATION = ARXIV:1503.07148;%%.

\bibitem{Ishii:2015gia}
T.~Ishii, E.~Kiritsis, and C.~Rosen, ``{Thermalization in a Holographic
  Confining Gauge Theory},''
  \href{http://dx.doi.org/10.1007/JHEP08(2015)008}{{\em JHEP} {\bf 08} (2015)
  008},
\href{http://arxiv.org/abs/1503.07766}{{\tt arXiv:1503.07766 [hep-th]}}.
%%CITATION = ARXIV:1503.07766;%%.

\bibitem{Attems:2016ugt}
M.~Attems, J.~Casalderrey-Solana, D.~Mateos, I.~Papadimitriou,
  D.~Santos-Oliván, C.~F. Sopuerta, M.~Triana, and M.~Zilhão,
  ``{Thermodynamics, transport and relaxation in non-conformal theories},''
\href{http://arxiv.org/abs/1603.01254}{{\tt arXiv:1603.01254 [hep-th]}}.
%%CITATION = ARXIV:1603.01254;%%.

\bibitem{Attems:2016tby}
M.~Attems, J.~Casalderrey-Solana, D.~Mateos, D.~Santos-Oliván, C.~F. Sopuerta,
  M.~Triana, and M.~Zilhão, ``{Collisions in Non-conformal Theories:
  Hydrodynamization without Equilibration},''
\href{http://arxiv.org/abs/1604.06439}{{\tt arXiv:1604.06439 [hep-th]}}.
%%CITATION = ARXIV:1604.06439;%%.

\bibitem{Buchel:2015ofa}
A.~Buchel and A.~Day, ``{Universal relaxation in quark-gluon plasma at strong
  coupling},'' \href{http://dx.doi.org/10.1103/PhysRevD.92.026009}{{\em Phys.
  Rev.} {\bf D92} (2015) no.~2, 026009},
\href{http://arxiv.org/abs/1505.05012}{{\tt arXiv:1505.05012 [hep-th]}}.
%%CITATION = ARXIV:1505.05012;%%.

\bibitem{Janik:2008tc}
R.~A. Janik and P.~Witaszczyk, ``{Towards the description of anisotropic plasma
  at strong coupling},''
  \href{http://dx.doi.org/10.1088/1126-6708/2008/09/026}{{\em JHEP} {\bf 09}
  (2008)  026},
\href{http://arxiv.org/abs/0806.2141}{{\tt arXiv:0806.2141 [hep-th]}}.
%%CITATION = ARXIV:0806.2141;%%.

\bibitem{Heinz:2004pj}
U.~W. Heinz, ``{Thermalization at RHIC},''
  \href{http://dx.doi.org/10.1063/1.1843595}{{\em AIP Conf.Proc.} {\bf 739}
  (2005)  163--180},
\href{http://arxiv.org/abs/nucl-th/0407067}{{\tt arXiv:nucl-th/0407067
  [nucl-th]}}.
%%CITATION = NUCL-TH/0407067;%%.

\bibitem{Chesler:2015fpa}
P.~M. Chesler, N.~Kilbertus, and W.~van~der Schee, ``{Universal hydrodynamic
  flow in holographic planar shock collisions},''
  \href{http://dx.doi.org/10.1007/JHEP11(2015)135}{{\em JHEP} {\bf 11} (2015)
  135},
\href{http://arxiv.org/abs/1507.02548}{{\tt arXiv:1507.02548 [hep-th]}}.
%%CITATION = ARXIV:1507.02548;%%.

\bibitem{Casalderrey-Solana:2016xfq}
J.~Casalderrey-Solana, D.~Mateos, W.~van~der Schee, and M.~Triana,
  ``{Holographic heavy ion collisions with baryon charge},''
  \href{http://dx.doi.org/10.1007/JHEP09(2016)108}{{\em JHEP} {\bf 09} (2016)
  108},
\href{http://arxiv.org/abs/1607.05273}{{\tt arXiv:1607.05273 [hep-th]}}.
%%CITATION = ARXIV:1607.05273;%%.

\bibitem{Ford:1999qv}
L.~H. Ford and T.~A. Roman, ``{The Quantum interest conjecture},''
  \href{http://dx.doi.org/10.1103/PhysRevD.60.104018}{{\em Phys. Rev.} {\bf
  D60} (1999)  104018},
\href{http://arxiv.org/abs/gr-qc/9901074}{{\tt arXiv:gr-qc/9901074 [gr-qc]}}.
%%CITATION = GR-QC/9901074;%%.

\bibitem{Arnold:2014jva}
P.~Arnold, P.~Romatschke, and W.~van~der Schee, ``{Absence of a local rest
  frame in far from equilibrium quantum matter},''
  \href{http://dx.doi.org/10.1007/JHEP10(2014)110}{{\em JHEP} {\bf 10} (2014)
  110},
\href{http://arxiv.org/abs/1408.2518}{{\tt arXiv:1408.2518 [hep-th]}}.
%%CITATION = ARXIV:1408.2518;%%.

\bibitem{Shen:2015msa}
C.~Shen and U.~Heinz, ``{The road to precision: Extraction of the specific
  shear viscosity of the quark-gluon plasma},''
  \href{http://dx.doi.org/10.1080/10619127.2015.1006502}{{\em Nucl. Phys. News}
  {\bf 25} (2015) no.~2, 6--11},
\href{http://arxiv.org/abs/1507.01558}{{\tt arXiv:1507.01558 [nucl-th]}}.
%%CITATION = ARXIV:1507.01558;%%.

\bibitem{Buchel:2008vz}
A.~Buchel, R.~C. Myers, and A.~Sinha, ``{Beyond eta/s = 1/4 pi},''
  \href{http://dx.doi.org/10.1088/1126-6708/2009/03/084}{{\em JHEP} {\bf 03}
  (2009)  084},
\href{http://arxiv.org/abs/0812.2521}{{\tt arXiv:0812.2521 [hep-th]}}.
%%CITATION = ARXIV:0812.2521;%%.

\bibitem{Buchel:2003tz}
A.~Buchel and J.~T. Liu, ``{Universality of the shear viscosity in
  supergravity},'' \href{http://dx.doi.org/10.1103/PhysRevLett.93.090602}{{\em
  Phys. Rev. Lett.} {\bf 93} (2004)  090602},
\href{http://arxiv.org/abs/hep-th/0311175}{{\tt arXiv:hep-th/0311175
  [hep-th]}}.
%%CITATION = HEP-TH/0311175;%%.

\bibitem{Janik:2006gp}
R.~A. Janik and R.~B. Peschanski, ``{Gauge/gravity duality and thermalization
  of a boost-invariant perfect fluid},''
  \href{http://dx.doi.org/10.1103/PhysRevD.74.046007}{{\em Phys. Rev.} {\bf
  D74} (2006)  046007},
\href{http://arxiv.org/abs/hep-th/0606149}{{\tt arXiv:hep-th/0606149
  [hep-th]}}.
%%CITATION = HEP-TH/0606149;%%.

\bibitem{Heller:2014wfa}
M.~P. Heller, R.~A. Janik, M.~Spaliński, and P.~Witaszczyk, ``{Coupling
  hydrodynamics to nonequilibrium degrees of freedom in strongly interacting
  quark-gluon plasma},''
  \href{http://dx.doi.org/10.1103/PhysRevLett.113.261601}{{\em Phys.Rev.Lett.}
  {\bf 113} (2014) no.~26, 261601},
\href{http://arxiv.org/abs/1409.5087}{{\tt arXiv:1409.5087 [hep-th]}}.
%%CITATION = ARXIV:1409.5087;%%.

\bibitem{Alver:2010dn}
B.~H. Alver, C.~Gombeaud, M.~Luzum, and J.-Y. Ollitrault, ``{Triangular flow in
  hydrodynamics and transport theory},''
  \href{http://dx.doi.org/10.1103/PhysRevC.82.034913}{{\em Phys. Rev.} {\bf
  C82} (2010)  034913},
\href{http://arxiv.org/abs/1007.5469}{{\tt arXiv:1007.5469 [nucl-th]}}.
%%CITATION = ARXIV:1007.5469;%%.

\bibitem{Bozek:2011if}
P.~Bozek, ``{Collective flow in p-Pb and d-Pd collisions at TeV energies},''
  \href{http://dx.doi.org/10.1103/PhysRevC.85.014911}{{\em Phys. Rev.} {\bf
  C85} (2012)  014911},
\href{http://arxiv.org/abs/1112.0915}{{\tt arXiv:1112.0915 [hep-ph]}}.
%%CITATION = ARXIV:1112.0915;%%.

\bibitem{Werner:2013tya}
K.~Werner, B.~Guiot, I.~Karpenko, and T.~Pierog, ``{Analysing radial flow
  features in p-Pb and p-p collisions at several TeV by studying identified
  particle production in EPOS3},''
  \href{http://dx.doi.org/10.1103/PhysRevC.89.064903}{{\em Phys. Rev.} {\bf
  C89} (2014) no.~6, 064903},
\href{http://arxiv.org/abs/1312.1233}{{\tt arXiv:1312.1233 [nucl-th]}}.
%%CITATION = ARXIV:1312.1233;%%.

\bibitem{Werner:2014xoa}
K.~Werner, B.~Guiot, I.~Karpenko, and T.~Pierog, ``{A unified description of
  the reaction dynamics from pp to pA to AA collisions},''
  \href{http://dx.doi.org/10.1016/j.nuclphysa.2014.08.093}{{\em Nucl. Phys.}
  {\bf A931} (2014)  83--91},
\href{http://arxiv.org/abs/1411.1048}{{\tt arXiv:1411.1048 [nucl-th]}}.
%%CITATION = ARXIV:1411.1048;%%.

\bibitem{Aniceto:2011nu}
I.~Aniceto, R.~Schiappa, and M.~Vonk, ``{The Resurgence of Instantons in String
  Theory},'' \href{http://dx.doi.org/10.4310/CNTP.2012.v6.n2.a3}{{\em
  Commun.Num.Theor.Phys.} {\bf 6} (2012)  339--496},
\href{http://arxiv.org/abs/1106.5922}{{\tt arXiv:1106.5922 [hep-th]}}.
%%CITATION = ARXIV:1106.5922;%%.

\bibitem{Argyres:2012vv}
P.~Argyres and M.~Unsal, ``{A semiclassical realization of infrared
  renormalons},'' \href{http://dx.doi.org/10.1103/PhysRevLett.109.121601}{{\em
  Phys.Rev.Lett.} {\bf 109} (2012)  121601},
\href{http://arxiv.org/abs/1204.1661}{{\tt arXiv:1204.1661 [hep-th]}}.
%%CITATION = ARXIV:1204.1661;%%.

\bibitem{Argyres:2012ka}
P.~C. Argyres and M.~Unsal, ``{The semi-classical expansion and resurgence in
  gauge theories: new perturbative, instanton, bion, and renormalon effects},''
  \href{http://dx.doi.org/10.1007/JHEP08(2012)063}{{\em JHEP} {\bf 1208} (2012)
   063},
\href{http://arxiv.org/abs/1206.1890}{{\tt arXiv:1206.1890 [hep-th]}}.
%%CITATION = ARXIV:1206.1890;%%.

\bibitem{Dunne:2012ae}
G.~V. Dunne and M.~Unsal, ``{Resurgence and Trans-series in Quantum Field
  Theory: The CP(N-1) Model},''
  \href{http://dx.doi.org/10.1007/JHEP11(2012)170}{{\em JHEP} {\bf 1211} (2012)
   170},
\href{http://arxiv.org/abs/1210.2423}{{\tt arXiv:1210.2423 [hep-th]}}.
%%CITATION = ARXIV:1210.2423;%%.

\bibitem{Aniceto:2013fka}
I.~Aniceto and R.~Schiappa, ``{Nonperturbative Ambiguities and the Reality of
  Resurgent Transseries},''
  \href{http://dx.doi.org/10.1007/s00220-014-2165-z}{{\em Commun.Math.Phys.}
  {\bf 335} (2015) no.~1, 183--245},
\href{http://arxiv.org/abs/1308.1115}{{\tt arXiv:1308.1115 [hep-th]}}.
%%CITATION = ARXIV:1308.1115;%%.

\bibitem{Cherman:2014ofa}
A.~Cherman, D.~Dorigoni, and M.~Unsal, ``{Decoding perturbation theory using
  resurgence: Stokes phenomena, new saddle points and Lefschetz thimbles},''
\href{http://arxiv.org/abs/1403.1277}{{\tt arXiv:1403.1277 [hep-th]}}.
%%CITATION = ARXIV:1403.1277;%%.

\bibitem{Denicol:2016bjh}
G.~S. Denicol and J.~Noronha, ``{Divergence of the Chapman-Enskog expansion in
  relativistic kinetic theory},''
\href{http://arxiv.org/abs/1608.07869}{{\tt arXiv:1608.07869 [nucl-th]}}.
%%CITATION = ARXIV:1608.07869;%%.

\bibitem{Son:2009tf}
D.~T. Son and P.~Surowka, ``{Hydrodynamics with Triangle Anomalies},''
  \href{http://dx.doi.org/10.1103/PhysRevLett.103.191601}{{\em Phys. Rev.
  Lett.} {\bf 103} (2009)  191601},
\href{http://arxiv.org/abs/0906.5044}{{\tt arXiv:0906.5044 [hep-th]}}.
%%CITATION = ARXIV:0906.5044;%%.

\bibitem{Huang:2011dc}
X.-G. Huang, A.~Sedrakian, and D.~H. Rischke, ``{Kubo formulae for relativistic
  fluids in strong magnetic fields},''
  \href{http://dx.doi.org/10.1016/j.aop.2011.08.001}{{\em Annals Phys.} {\bf
  326} (2011)  3075--3094},
\href{http://arxiv.org/abs/1108.0602}{{\tt arXiv:1108.0602 [astro-ph.HE]}}.
%%CITATION = ARXIV:1108.0602;%%.

\bibitem{Basar:2015ava}
G.~Basar and G.~V. Dunne, ``{Hydrodynamics, resurgence, and
  transasymptotics},'' \href{http://dx.doi.org/10.1103/PhysRevD.92.125011}{{\em
  Phys. Rev.} {\bf D92} (2015) no.~12, 125011},
\href{http://arxiv.org/abs/1509.05046}{{\tt arXiv:1509.05046 [hep-th]}}.
%%CITATION = ARXIV:1509.05046;%%.

\bibitem{Aniceto:2015mto}
I.~Aniceto and M.~Spaliński, ``{Resurgence in Extended Hydrodynamics},''
  \href{http://dx.doi.org/10.1103/PhysRevD.93.085008}{{\em Phys. Rev.} {\bf
  D93} (2016) no.~8, 085008},
\href{http://arxiv.org/abs/1511.06358}{{\tt arXiv:1511.06358 [hep-th]}}.
%%CITATION = ARXIV:1511.06358;%%.

\bibitem{Warnick:2013hba}
C.~M. Warnick, ``{On quasinormal modes of asymptotically anti-de Sitter black
  holes},'' \href{http://dx.doi.org/10.1007/s00220-014-2171-1}{{\em Commun.
  Math. Phys.} {\bf 333} (2015) no.~2, 959--1035},
\href{http://arxiv.org/abs/1306.5760}{{\tt arXiv:1306.5760 [gr-qc]}}.
%%CITATION = ARXIV:1306.5760;%%.

\bibitem{Grozdanov:2016vgg}
S.~Grozdanov, N.~Kaplis, and A.~O. Starinets, ``{From strong to weak coupling
  in holographic models of thermalization},''
  \href{http://dx.doi.org/10.1007/JHEP07(2016)151}{{\em JHEP} {\bf 07} (2016)
  151},
\href{http://arxiv.org/abs/1605.02173}{{\tt arXiv:1605.02173 [hep-th]}}.
%%CITATION = ARXIV:1605.02173;%%.

\bibitem{Grozdanov:2016zjj}
S.~Grozdanov and W.~van~der Schee, ``{Coupling constant corrections in
  holographic heavy ion collisions},''
\href{http://arxiv.org/abs/1610.08976}{{\tt arXiv:1610.08976 [hep-th]}}.
%%CITATION = ARXIV:1610.08976;%%.

\bibitem{Andrade:2016rln}
T.~Andrade, J.~Casalderrey-Solana, and A.~Ficnar, ``{Holographic Isotropisation
  in Gauss-Bonnet Gravity},''
\href{http://arxiv.org/abs/1610.08987}{{\tt arXiv:1610.08987 [hep-th]}}.
%%CITATION = ARXIV:1610.08987;%%.

\bibitem{Balasubramanian:2011ur}
V.~Balasubramanian, A.~Bernamonti, J.~de~Boer, N.~Copland, B.~Craps,
  E.~Keski-Vakkuri, B.~Muller, A.~Schafer, M.~Shigemori, and W.~Staessens,
  ``{Holographic Thermalization},''
  \href{http://dx.doi.org/10.1103/PhysRevD.84.026010}{{\em Phys. Rev.} {\bf
  D84} (2011)  026010},
\href{http://arxiv.org/abs/1103.2683}{{\tt arXiv:1103.2683 [hep-th]}}.
%%CITATION = ARXIV:1103.2683;%%.

\bibitem{Ecker:2016thn}
C.~Ecker, D.~Grumiller, P.~Stanzer, S.~A. Stricker, and W.~van~der Schee,
  ``{Exploring nonlocal observables in shock wave collisions},''
\href{http://arxiv.org/abs/1609.03676}{{\tt arXiv:1609.03676 [hep-th]}}.
%%CITATION = ARXIV:1609.03676;%%.

\bibitem{CaronHuot:2011dr}
S.~Caron-Huot, P.~M. Chesler, and D.~Teaney, ``{Fluctuation, dissipation, and
  thermalization in non-equilibrium AdS$_5$ black hole geometries},''
  \href{http://dx.doi.org/10.1103/PhysRevD.84.026012}{{\em Phys. Rev.} {\bf
  D84} (2011)  026012},
\href{http://arxiv.org/abs/1102.1073}{{\tt arXiv:1102.1073 [hep-th]}}.
%%CITATION = ARXIV:1102.1073;%%.

\bibitem{Chesler:2011ds}
P.~M. Chesler and D.~Teaney, ``{Dynamical Hawking Radiation and Holographic
  Thermalization},''
\href{http://arxiv.org/abs/1112.6196}{{\tt arXiv:1112.6196 [hep-th]}}.
%%CITATION = ARXIV:1112.6196;%%.

\bibitem{Chesler:2012zk}
P.~M. Chesler and D.~Teaney, ``{Dilaton emission and absorption from
  far-from-equilibrium non-abelian plasma},''
\href{http://arxiv.org/abs/1211.0343}{{\tt arXiv:1211.0343 [hep-th]}}.
%%CITATION = ARXIV:1211.0343;%%.

\bibitem{Keranen:2014lna}
V.~Keranen and P.~Kleinert, ``{Non-equilibrium scalar two point functions in
  AdS/CFT},'' \href{http://dx.doi.org/10.1007/JHEP04(2015)119}{{\em JHEP} {\bf
  04} (2015)  119},
\href{http://arxiv.org/abs/1412.2806}{{\tt arXiv:1412.2806 [hep-th]}}.
%%CITATION = ARXIV:1412.2806;%%.

\bibitem{Keranen:2015mqc}
V.~Keranen and P.~Kleinert, ``{Thermalization of Wightman functions in AdS/CFT
  and quasinormal modes},''
  \href{http://dx.doi.org/10.1103/PhysRevD.94.026010}{{\em Phys. Rev.} {\bf
  D94} (2016) no.~2, 026010},
\href{http://arxiv.org/abs/1511.08187}{{\tt arXiv:1511.08187 [hep-th]}}.
%%CITATION = ARXIV:1511.08187;%%.

\bibitem{David:2015xqa}
J.~R. David and S.~Khetrapal, ``{Thermalization of Green functions and
  quasinormal modes},'' \href{http://dx.doi.org/10.1007/JHEP07(2015)041}{{\em
  JHEP} {\bf 07} (2015)  041},
\href{http://arxiv.org/abs/1504.04439}{{\tt arXiv:1504.04439 [hep-th]}}.
%%CITATION = ARXIV:1504.04439;%%.

\bibitem{Bizon:2011gg}
P.~Bizon and A.~Rostworowski, ``{On weakly turbulent instability of anti-de
  Sitter space},'' \href{http://dx.doi.org/10.1103/PhysRevLett.107.031102}{{\em
  Phys. Rev. Lett.} {\bf 107} (2011)  031102},
\href{http://arxiv.org/abs/1104.3702}{{\tt arXiv:1104.3702 [gr-qc]}}.
%%CITATION = ARXIV:1104.3702;%%.

\bibitem{Maliborski:2013jca}
M.~Maliborski and A.~Rostworowski, ``{Time-Periodic Solutions in an Einstein
  AdS–Massless-Scalar-Field System},''
  \href{http://dx.doi.org/10.1103/PhysRevLett.111.051102}{{\em Phys. Rev.
  Lett.} {\bf 111} (2013)  051102},
\href{http://arxiv.org/abs/1303.3186}{{\tt arXiv:1303.3186 [gr-qc]}}.
%%CITATION = ARXIV:1303.3186;%%.

\bibitem{Balasubramanian:2014cja}
V.~Balasubramanian, A.~Buchel, S.~R. Green, L.~Lehner, and S.~L. Liebling,
  ``{Holographic Thermalization, Stability of Anti–de Sitter Space, and the
  Fermi-Pasta-Ulam Paradox},''
  \href{http://dx.doi.org/10.1103/PhysRevLett.113.071601}{{\em Phys. Rev.
  Lett.} {\bf 113} (2014) no.~7, 071601},
\href{http://arxiv.org/abs/1403.6471}{{\tt arXiv:1403.6471 [hep-th]}}.
%%CITATION = ARXIV:1403.6471;%%.

\bibitem{Craps:2015jma}
B.~Craps and O.~Evnin, ``{AdS (in)stability: an analytic approach},''
  \href{http://dx.doi.org/10.1002/prop.201500067}{{\em Fortsch. Phys.} {\bf 64}
  (2016)  336--344},
\href{http://arxiv.org/abs/1510.07836}{{\tt arXiv:1510.07836 [gr-qc]}}.
%%CITATION = ARXIV:1510.07836;%%.

\bibitem{Carrasco:2012nf}
F.~Carrasco, L.~Lehner, R.~C. Myers, O.~Reula, and A.~Singh, ``{Turbulent flows
  for relativistic conformal fluids in 2+1 dimensions},''
  \href{http://dx.doi.org/10.1103/PhysRevD.86.126006}{{\em Phys. Rev.} {\bf
  D86} (2012)  126006},
\href{http://arxiv.org/abs/1210.6702}{{\tt arXiv:1210.6702 [hep-th]}}.
%%CITATION = ARXIV:1210.6702;%%.

\bibitem{Adams:2013vsa}
A.~Adams, P.~M. Chesler, and H.~Liu, ``{Holographic turbulence},''
  \href{http://dx.doi.org/10.1103/PhysRevLett.112.151602}{{\em Phys. Rev.
  Lett.} {\bf 112} (2014) no.~15, 151602},
\href{http://arxiv.org/abs/1307.7267}{{\tt arXiv:1307.7267 [hep-th]}}.
%%CITATION = ARXIV:1307.7267;%%.

\bibitem{Green:2013zba}
S.~R. Green, F.~Carrasco, and L.~Lehner, ``{Holographic Path to the Turbulent
  Side of Gravity},'' \href{http://dx.doi.org/10.1103/PhysRevX.4.011001}{{\em
  Phys. Rev.} {\bf X4} (2014) no.~1, 011001},
\href{http://arxiv.org/abs/1309.7940}{{\tt arXiv:1309.7940 [hep-th]}}.
%%CITATION = ARXIV:1309.7940;%%.

\bibitem{Yang:2014tla}
H.~Yang, A.~Zimmerman, and L.~Lehner, ``{Turbulent Black Holes},''
  \href{http://dx.doi.org/10.1103/PhysRevLett.114.081101}{{\em Phys. Rev.
  Lett.} {\bf 114} (2015)  081101},
\href{http://arxiv.org/abs/1402.4859}{{\tt arXiv:1402.4859 [gr-qc]}}.
%%CITATION = ARXIV:1402.4859;%%.

\end{thebibliography}\endgroup
\bibliographystyle{utphys}

\end{document}